\documentclass[useAMS,usenatbib,usegraphicx]{mn2e}

\usepackage{amsmath,amssymb,amsbsy,amsfonts}
\usepackage[british]{babel}
\usepackage{mdwmath}
\usepackage{color}
\usepackage{url}
\usepackage{mathrsfs}
\usepackage{array}
\usepackage{graphicx}
\usepackage{units}
\usepackage{hyperref}
\usepackage{times}
\usepackage{pdflscape}
\usepackage{balance}
\definecolor{custom-blue}{RGB}{41,22,206}
\hypersetup{
  colorlinks   = true, %Colours links instead of ugly boxes
  urlcolor     = custom-blue, %Colour for external hyperlinks
  linkcolor    = custom-blue, %Colour of internal links
  citecolor    = custom-blue  %Colour of citations
}

\usepackage[abs]{overpic}
\usepackage{xcolor,varwidth}

\newcommand{\eq}[1]{equation~\eqref{#1}}
\newcommand{\eqs}[1]{equations~\eqref{#1}}
\newcommand{\cc}{\mathrm{c}}
\newcommand{\co}{\mathrm{const.}}
\newcommand{\ud}{\mathrm{d}}
\newcommand{\G}{\mathrm{G}}
\newcommand{\s}[1]{{\text{\tiny $#1$ }}\hspace{-2pt}}

\newcommand{\8}{\s{\infty}}

\renewcommand{\theequation}{\thesection.\arabic{equation}}

\title[Choked accretion]{Choked accretion: 
from radial infall to bipolar outflows by breaking spherical symmetry}

\author[A.~Aguayo-Ortiz, E.~Tejeda \& X.~Hernandez]
       {Alejandro Aguayo-Ortiz$^1$, Emilio Tejeda$^{2}$ and X. 
Hernandez$^1$\thanks{E-mail: aaguayo@astro.unam.mx;
emilio.tejeda@conacyt.mx; xavier@astro.unam.mx}  \\
$^1$ Instituto de Astronom\'ia, Universidad Nacional Aut\'onoma de M\'exico, AP 
70-264, 04510 Ciudad de M\'exico, Mexico \\
$^2$ C\'atedras CONACyT --  Instituto de F\'isica y Matem\'aticas, Universidad Michoacana 
de San Nicol\'as de Hidalgo, Edificio C-3, Ciudad Universitaria, 58040 
Morelia,\\
 \hspace{0.15cm} Michoac\'an, Mexico}

\pagerange{\pageref{firstpage}--\pageref{lastpage}}

\begin{document}

\label{firstpage}

\maketitle

\begin{abstract}

Steady state, spherically symmetric accretion flows are well understood
in terms of the Bondi solution. Spherical symmetry however, is necessarily
an idealized approximation to reality. Here we explore the consequences of
deviations away from spherical symmetry, first through a simple analytic model
to motivate the physical processes involved, and then through hydrodynamical,
numerical simulations of an ideal fluid accreting onto a Newtonian gravitating
object. Specifically, we consider axisymmetric, large-scale, small amplitude
deviations in the density field such that the equatorial plane is over dense
as compared to the polar regions. We find that the resulting polar density
gradient dramatically alters the Bondi result and gives rise to steady state
solutions presenting bipolar outflows. As the density contrast increases, more
and more material is ejected from the system, attaining speeds larger than
the local escape velocities for even modest density contrasts. Interestingly,
interior to the outflow region, the flow tends locally towards the Bondi
solution, with a resulting total mass accretion rate through the inner
boundary {\it choking} at a value very close to the corresponding Bondi
one. Thus, the numerical experiments performed suggest the appearance of a
maximum achievable accretion rate, with any extra material being ejected,
even for very small departures from spherical symmetry.

\end{abstract}

\begin{keywords}
  accretion, accretion discs -- gravitation -- hydrodynamics -- methods: 
numerical.
\end{keywords}

\section{Introduction} 
\label{S1}

The accretion of fluids towards gravitational objects is a topic of general
interest in astrophysics, as such phenomena underpin the physics of large
classes of systems \citep{hawley2015}. Notably, the ubiquitous jets observed
from Young Stellar Objects (YSOs) to Gamma Ray Bursts (GRBs) and Active
Galactic Nuclei (AGN), are fuelled by the accretion of gas on to massive
objects. Although the detailed physics of accretion problems is complex,
including the effects of rotation, magnetic fields, non-ideal fluids and
mixing, to name but a few~\citep{pudritz2007,tchekhovskoy2015,jafari2019},
the availability of simplified analytic solutions where the salient physical
ingredients can be transparently traced, has always provided valuable insights
and well understood limiting cases for the analysis of these systems.

The first analytic solution to such an accretion problem was the Newtonian
spherically symmetric model of \cite{bondi52}, with the corresponding extension
to general relativity by \cite{michel72}. In both cases, the authors assumed
stationariness, spherical symmetry and zero angular momentum for the infalling
material, which was in turn described as an ideal fluid.

On the other hand, the origin of jets is typically understood in connection
with accretion disc models, where the geometry very strongly deviates
from spherical symmetry. In these, rotation and small-scale magnetic
fields are considered as fundamental for the stability and evolution of
the disc \citep{shakura,balbus}, while the interplay of these with a
large-scale magnetic field is thought to be responsible for the launching
and subsequent collimation of the jet~\citep{hawley2015}. This so-called
magneto-rotational mechanism has been studied both at a Newtonian level
\citep{lovelace76, blandford76, BP1982}  and in general relativity when a
central black hole is involved \citep{BZ1977}.

The viability of the magneto-rotational mechanism for launching powerful jets
has been successfully demonstrated by means of magneto-hydrodynamic (MHD)
numerical simulations. This has been done at a non-relativistic level for
jets associated to YSOs~\citep{casse2002}, as well as with comprehensive,
general relativistic-MHD simulations of accretion discs around spinning
black holes~\citep{semenov04, qian2018,sheikhnezami2018,liska2019}.

The similarity of jets across a vast range of astrophysical scales, from
quasars to micro-quasars \citep[e.g.][]{MR94}, along with open questions
regarding the matter content of relativistic jets~\citep{hawley2015}
or the link between the accretion disc and the acceleration
process~\citep{romero2017}, might hint towards the presence in some cases
of more simple outflow-producing mechanisms based on hydrodynamical
physics. Moreover, it is clear that if some kind of universal mechanism
underlies all astrophysical jets, then it can not rely on the central accretor
being a black hole.

Hydrodynamical models have been introduced for collimating and accelerating
jets, in the context of AGNs \citep{BR74} and of YSOs \citep[specifically H-H
objects as studied by][]{CR80}. On the other hand, a purely hydrodynamical
mechanism, where a small amplitude density gradient on the inflow boundary
conditions yields and inflow/outflow steady state solution, was introduced
by \cite{hernandez14} based on an analytic perturbation analysis of the
hydrodynamic equations for an isothermal fluid.

In \cite{hernandez14}, the authors consider an originally radial
accretion flow that becomes increasingly dense as the fluid approaches the
central accretor. As the authors assume that this flow originates from
the inner walls of a disc-like configuration, there is a certain degree of
inhomogeneity in the density field, with the polar regions being less dense
than the disc plane. As a result of this density inhomogeneity, on approaching
the central regions the geometrical focusing of the accreted material results
in a pressure gradient that deviates some of the fluid elements from their
infall trajectories and expels them from the central region along a bipolar
outflow.\footnote{A similar deviation process by a pressure gradient is
incorporated in the circulation model presented by \citet{lery} in the context
of molecular outflows in YSOs.} Within this scenario, the resulting bipolar
outflow is neither accelerated to large Mach numbers, nor collimated as a
proper jet. 

Following on from \citet{hernandez14}, in this work we extend the previous
results by considering more realistic adiabatic indices for the infalling
material. The problem can no longer be treated analytically, so we implement
full hydrodynamical numerical simulations using the free GPL hydrodynamical
code {\it aztekas} \citep{OM08,aguayo18}.\footnote{\url{aztekas.org}
\copyright 2008 Sergio Mendoza \& Daniel Olvera and \copyright 2018
Alejandro Aguayo-Ortiz \& Sergio Mendoza. The code can be downloaded from
\url{github.com/aztekas-code/aztekas-main}.}

We have found with these simulations a flux-limited accretion mode, in which
the total mass infall rate  onto the central accretor is limited by a fixed
value.  This value coincides very closely with the mass accretion rate of the
spherically symmetric Bondi solution. Whenever the incoming accretion flow
surpasses this threshold value, the excess flow is redirected by a density
gradient and expelled through the poles. Since the incoming accretion flow
is jamming at a gravitational bottleneck, we refer to this ejection mechanism
as {\it choked accretion}.

With the numerical simulations presented in this work we recover the main
result of \citet{hernandez14} that a large scale, small amplitude
inhomogeneity in the density field can lead to the onset of a bipolar
outflow as a generic result. The choked accretion model can then be viewed
as a transition bridge between the spherically symmetric condition treated
by Bondi and Michel, where no outflows appear, and the disc geometries of
the jet-generating models mentioned above.

In a separate work,~\citet{tah2019}, we are proposing a full-analytic,
general relativistic model of the choked accretion mechanism.  This analytic
model describes an ultra relativistic gas with a stiff equation of state
\citep[see][]{petrich88} and is based on the conditions of steady state,
axisymmetry and irrotational flow. As shown by \cite{tejeda18}, the
non-relativistic limit of such a model corresponds to an incompressible
fluid in Newtonian hydrodynamics. With this motivation in mind, we present
in this article a simple analytic model of an incompressible fluid.

The remainder of the paper is organized as follows. In Section~\ref{S2},
we present a simple analytic model of the choked accretion that leads to an
inflow/outflow configuration. Section~\ref{S3} presents the hydrodynamical
simulations used to extend this model to more general conditions, allowing
for a range of plausible adiabatic indices. The code is validated and tested
in the spherically symmetric case, where steady state solutions accurately
tracing the corresponding Bondi ones are recovered. In Section~\ref{S4}
we analyse the results obtained and discuss the applicability of the choked
accretion mechanism in astrophysical settings. Finally, in Section~\ref{S5}
we present our conclusions.

\setcounter{equation}{0}
\section{Analytic model}
\label{S2}

In order to present a simple model where the physics leading from the
breakage of spherical symmetry to the establishing of an outflow can be traced
transparently, in this section we discuss an analytic model of choked accretion.
Following \cite{tejeda18}, this model is based on an incompressible fluid
under the approximations of steady state, axisymmetry and irrotational flow.

This model can be considered as complementary to the perturbation
analysis presented in \cite{hernandez14},
where isothermal conditions were assumed, and which shows that a
spherically symmetric flow onto a point mass is in fact unstable towards the
development of the outflow phenomenology discussed here, as soon as a slight
perturbation to spherical symmetry is introduced in the inflow conditions. 

Under the irrotational flow condition, the fluid's velocity field
can be obtained as the gradient of a velocity potential $\Phi$
(i.e.~$\vec{v}=\nabla\Phi$). For an incompressible fluid, this means
that $\Phi$  has to be a solution to the Laplace equation $\nabla^2\Phi =
0$. Adopting spherical coordinates and imposing the axisymmetry condition,
we know that a general solution for $\Phi$ is given by \citep{currie}

\begin{equation}
\Phi = \sum_{n=0}^\8 \left(A_n\,r^n + B_n\,r^{-(n+1)} \right) 
P_n(\cos\theta),\label{ettt}
\end{equation}
where $A_n$ and $B_n$ are constant coefficients and $P_n(\cos\theta)$ 
is the Legendre polynomial of degree $n$. 

The lowest order solution that describes both accretion onto a central object
and an axisymmetric, bipolar outflow has as only non-vanishing coefficients
$B_0$ and $A_2$.\footnote{A scenario of wind accretion can be studied by
taking instead only $B_0$ and $A_1$ different from zero, as has been done by
\cite{petrich88} and \cite{tejeda18}. More complex geometries can be described
by considering higher order multipoles of $A_n$.}  Let us re-parametrize
these two coefficients as $B_0 = \alpha$, $A_2 = \alpha/(2\,\mathcal{S}^3)$,
and write the solution as
\begin{equation}
\Phi = \frac{\alpha}{r}\left[1 + \frac{r^3}{4\,\mathcal{S}^3}(3\cos^2\theta - 
1)\right],
\end{equation}
which leads to the velocity field
\begin{gather}
\frac{\ud r}{\ud t} =  - \frac{\alpha}{r^2} \left[1 - 
\frac{r^3}{2\,\mathcal{S}^3} 
\left(3\cos ^2\theta -1\right)\right] ,
\label{vr}\\
\frac{\ud \theta}{\ud t} =-\frac{3\, \alpha\, \sin\theta \, \cos \theta }{2\, 
\mathcal{S}^3}.
\label{vth}
\end{gather}

This velocity profile has the same dependence on the polar angle as the one
obtained for the isothermal hydrodynamic solutions of \citet{hernandez14},
which are also given in terms of Legendre polynomials.

\begin{figure}
\begin{center}
\includegraphics[width=\linewidth]{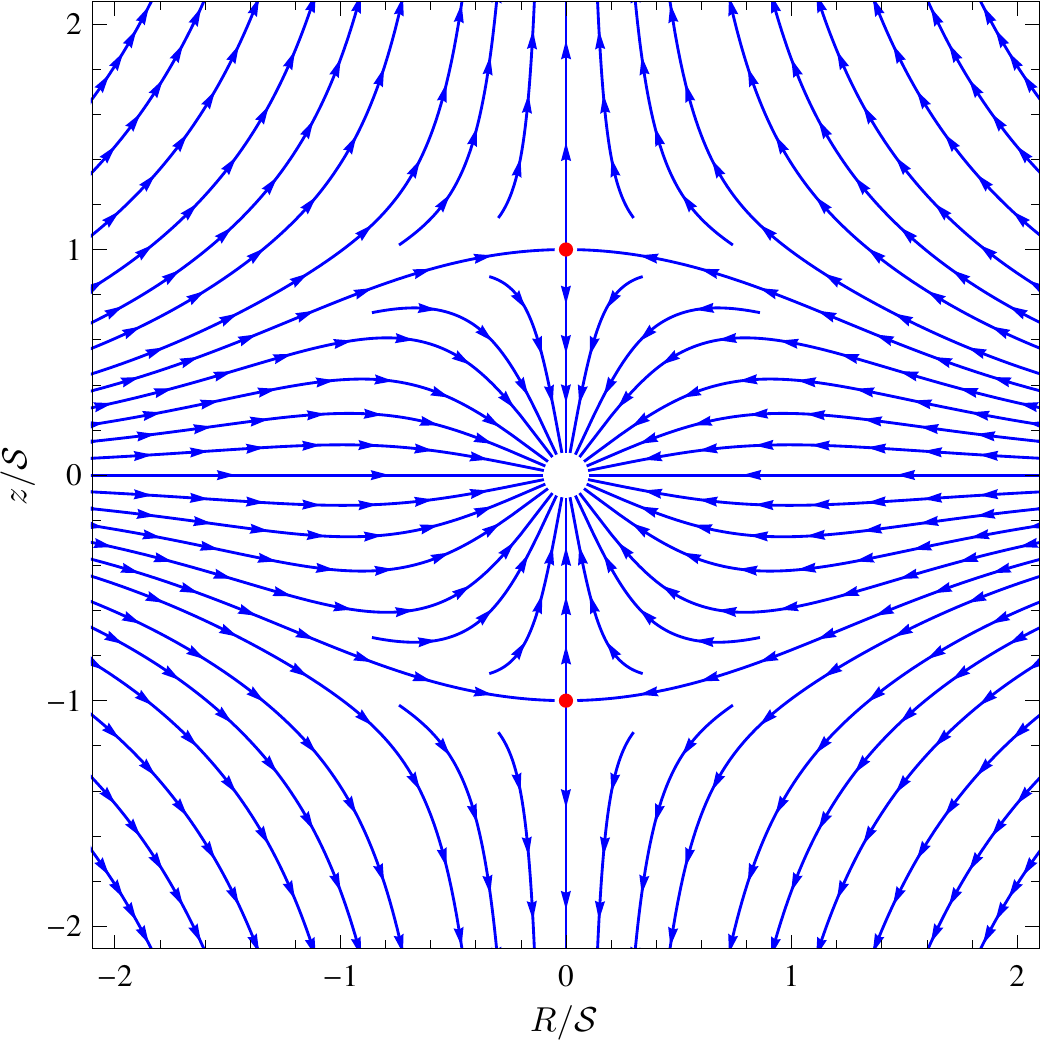} 
\end{center}
\caption{Streamlines of the incompressible analytic model showing an inner 
quasi-spherical accretion region, and polar stagnation points beyond which a 
polar outflow solution results. The red dots indicate the location of the
stagnation points along the symmetry axis.  The axes correspond to the usual 
cylindrical coordinates $R = r\,\sin\theta$, $z = r\,\cos\theta$.}
\label{fig1}
\end{figure}

The parameter $\mathcal{S}$ is related to the location of the stagnation point,
as from \eqs{vr} and \eqref{vth} we see that the points ($r=\mathcal{S}$,
$\theta = 0$) and ($r=\mathcal{S}$, $\theta = \pi$) correspond to the
stagnation points of the flow. In Figure~\ref{fig1} we show an example of
the resulting streamlines.

%---------------------------------------------------------------------
%---------------------------------------------------------------------

On the other hand, the parameter $\alpha$ is related to the total accretion 
rate onto the central object in the following way
\begin{equation}
\dot{M} = - 2\pi \int_0^\pi \rho\, \frac{\ud r}{\ud t} \,r^2 \sin\theta\, \ud 
\theta = 4\pi\rho\,\alpha ,
\label{e2.5}
\end{equation}
that is
\begin{equation}
 \alpha = \frac{\dot{M}}{4\pi\rho}.
\end{equation}

At this point $\dot{M}$ can have any arbitrary value. 
For the choked accretion model, we shall assume that $\dot{M}$
is given by the Bondi accretion rate, i.e.
\begin{equation}
 \alpha =\frac{\dot{M}_\mathrm{B}}{4\pi\rho} = \frac{1}{4}\frac{(\G M)^2}{a\8^3}
\left(\frac{2}{5-3\,\gamma}\right)^{\frac{5-3\,\gamma}{2(\gamma-1)} },
\label{Mbondi}
\end{equation}
where $M$ is the mass of the central accretor, $\gamma$ the adiabatic index of 
the fluid and $a\8$ the speed of sound far away from the central object. 

Let us consider that the flow is continuously being injected from a sphere of 
radius $\mathcal{R}$ that we shall refer to as the injection sphere. Provided that 
$\mathcal{S}<\mathcal{R}$, from \eq{vr} we see that the radial velocity changes 
sign at
\begin{equation}
 \cos\theta_0 = 
\sqrt{\frac{1}{3}\left(1+2\,\frac{\mathcal{S}^3}{\mathcal{R}^3}\right)}.
\end{equation}
More specifically, the flow is characterized by an inflow/outflow geometry, with 
inflow ($\dot{r}<0$) across the equatorial belt defined by $\theta_0 < \theta < 
\pi - \theta_0$ and outflow ($\dot{r}>0$) across the polar caps defined by 
$\theta <  \theta_0$ and $\theta > \pi - \theta_0$.

We can calculate now the mass injection rate across the sphere of radius 
$\mathcal{R}$ as
\begin{equation}
\begin{split}
 \dot{M}_\mathrm{in} & = - 4\pi \int_{\theta_0}^{\pi/2} \rho\, \frac{\ud r}{\ud 
t} 
\,\mathcal{R}^2 \sin\theta\, \ud \theta , \\
 & =  4\pi\rho\,\alpha \, 
\frac{\mathcal{R}^3}{\mathcal{S}^3}\left[\frac{1}{3}\left(1+2\,\frac{\mathcal{S}
^3}{\mathcal{R}^3}\right) 
\right]^{3/2}.
\end{split}
\label{e2.9}
\end{equation}
Similarly, we define the mass ejection rate leaving this same sphere as
\begin{equation}
\begin{split}
 \dot{M}_\mathrm{ej} & = 4\pi \int_{0}^{\theta_0} \rho\, \frac{\ud r}{\ud t} 
\,\mathcal{R}^2 \sin\theta\, \ud \theta , \\
 & =  \dot{M}_\mathrm{in} - \dot{M}.
\end{split}
\label{e2.10}
\end{equation}
Note that if $\mathcal{S}\ge \mathcal{R}$ then $\dot{M}_\mathrm{in} = \dot{M}$ 
and $\dot{M}_\mathrm{ej} = 0$. 

Finally, an expression for the streamlines can be found by combining \eqs{vr}
and \eqref{vth} as
\begin{equation}
\frac{\ud r}{\ud \theta} =    \frac{2\,\mathcal{S}^3 - r^3 
\left(3\cos ^2\theta -1\right)}{3\, r^2 \sin\theta \, \cos \theta }   ,
\end{equation}
which can be integrated to give
\begin{equation}
 r = S\left(2\frac{\Psi-\cos\theta}{\cos\theta\,\sin^2\theta} 
\right)^{1/3},
\end{equation}
where $\Psi$ is an integration constant (stream function). From this expression
we see that the streamlines corresponding to $\Psi = \pm1$ are the only
ones that arrive at the stagnation points located at $(\mathcal{S},\,0)$
and $(\mathcal{S},\,\pi)$, respectively. On the other hand, all of the
streamlines with $|\Psi| < 1$ accrete onto the central object while those
with $|\Psi| > 1$ escape along the bipolar outflow.

\setcounter{equation}{0}
\section{Numerical simulations}
\label{S3}

The simple toy model presented in the previous section is clearly limited 
by the assumption of an incompressible fluid. In this section we relax this 
condition and use instead full-hydrodynamic numerical simulations of a
polytropic fluid obeying
\begin{equation}
 P = K\,\rho^\gamma,
 \label{e3.0}
\end{equation}
where $P$ is the fluid pressure and $K=\co$ In this work we consider three 
different values for the adiabatic index, namely \mbox{$\gamma = 1$} 
corresponding to an isothermal fluid, $\gamma = 4/3$ describing a gas composed 
of relativistic particles,\footnote{This value is relevant not only for 
relativistic particles, but is also commonly used in astrophysical situations, 
e.g.~for optically thick material where the internal energy is dominated by 
radiation pressure~\citep[e.g.][]{awe2011}.} and $\gamma = 7/5$ corresponding to 
the adiabatic index of a diatomic gas.

In order to capture the basic premise on which the mechanism of choked 
accretion operates, i.e.~breaking spherical symmetry by introducing a small 
density contrast between the equator and the poles, here we parametrize this 
anisotropy by imposing the following density profile as boundary condition at 
the injection sphere ($r=\mathcal{R}$)
\begin{equation}
 \rho(\theta) = \rho_0 \left(1-\delta\cos^2\theta\right),
 \label{e3.1}
\end{equation} 
where $\rho_0$ is an arbitrary value that we set as the density unit and
$\delta$ is the density contrast between the equator and the poles defined as
\begin{equation}
 \delta = 1-\frac{\rho(0)}{\rho(\pi/2)}.
 \label{e3.2}
\end{equation}

We study this problem by means of numerical simulations performed
with the hydrodynamical code {\em aztekas}.  This code numerically solves
the inviscid Euler equations in a conservative form,
\begin{gather}
\frac{\partial \rho}{\partial t} + \nabla \cdot \left( \rho \mathbf{v} \right) =
0,
\label{eq:dens1} \\
\frac{\partial \left( \rho \mathbf{v} \right)}{\partial t} + \nabla \cdot \left(
\rho \mathbf{v} \otimes \mathbf{v}\right) + \nabla P = -\rho
\frac{\G M}{r^2} \hat{r},
\label{eq:momentum} \\
\frac{\partial E}{\partial t} + \nabla \cdot \left[ \mathbf{v} \left( 
E + P \right) \right] = -\rho \frac{\G M}{r^2} \mathbf{v} \cdot \hat{r},
\label{eq:energy1}
\end{gather}
with $\mathbf{v}$ the fluid velocity vector and $E$ the total energy density 
defined as
\begin{equation}
E = \frac{1}{2} \rho |\mathbf{v}|^2 + \epsilon,
\label{eq:total-energy}
\end{equation}
where $\epsilon$ is the internal energy density.
From \eq{e3.0}, together with the first law of 
thermodynamics for an ideal gas, we can close this system of equations with the 
following equation of state that relates the pressure to the internal energy
density
\begin{equation}
\epsilon = \frac{P}{\gamma - 1}.
\label{eq:eos}
\end{equation}

The \eqs{eq:dens1}-\eqref{eq:eos} are spatially discretized using a finite 
volume scheme together with a High Resolution Shock Capturing (HRSC) method, 
that uses a second order piecewise linear reconstructor (MC) and the 
HLL~\citep{hll1983} approximate Riemann solver to compute the numerical fluxes. 
The time evolution of these equations is calculated using a second order total 
variation diminishing Runge-Kutta time integrator (RK2)~\citep{shu1988}, with a 
Courant factor of 0.25.

Since the density profile in \eq{e3.1} is independent of the azimuthal angle, we 
assume that the problem retains symmetry with respect to  both the polar 
axis (axisymmetry) and the equatorial plane (north-south symmetry).  Hence, for 
the numerical simulations we adopt a two-dimensional spherical domain consisting 
of a uniform polar grid on 
$\theta\in[0,\pi/2]$  and a exponential radial grid with $r\in[\mathcal{R}_{\rm acc},\,\mathcal{R}]$, computed as 
\begin{equation}
 r_i = \mathcal{R}_{\rm acc} + 
 {\rm exp} \left( \frac{\ln (\mathcal{R} - \mathcal{R}_{\rm acc})i}{N_r} \right),
\label{e3.9}
\end{equation}
where $\mathcal{R}_{\rm acc}$ is the radius of the inner boundary. As fiducial
values for the numerical resolution we adopt  $N_r = N_\theta = 150$, while for
the radial boundaries we take $\mathcal{R}_{\rm acc} = 0.1\,r_\mathrm{B}$
and $\mathcal{R} = 10\,r_\mathrm{B}$, where $r_\mathrm{B} = \G M/a^2_\infty$
is the Bondi radius. The code uses $r_\mathrm{B}, \ a\8$ and $ t_\mathrm{B}
= r_\mathrm{B}/a\8$ as units of length, velocity and time, respectively.

%---------------------------------------------------------------------
\begin{figure}
\centering
\includegraphics[width=\linewidth]{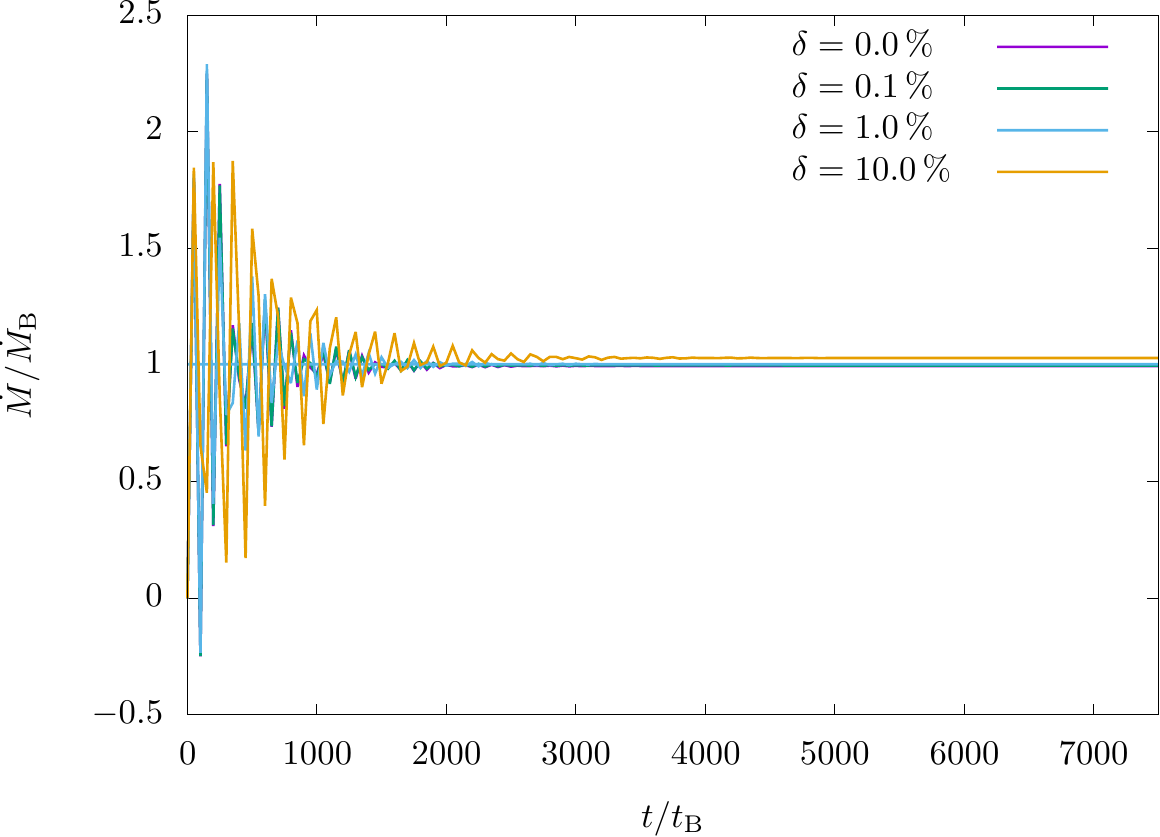}
\caption{ Mass accretion rate as a function of time for the four values of 
the density contrast $\delta$ in the case  $\gamma = 4/3$ (similar results were 
obtained for the other two values of $\gamma$). There is a clear convergence 
towards a steady state solution.} 
\label{fig2}
\end{figure}
%---------------------------------------------------------------------

%---------------------------------------------------------------------
\subsection{Initial and boundary conditions}

We performed different simulations using three values of the adiabatic
index $\gamma = 1, \, 4/3, \, 7/5,$ and four values of the density contrast
$\delta = 0, \, 0.1\%, \, 1\%$, and 10$\%$. As boundary conditions, we set
free-outflow\footnote{The free-outflow condition is implemented by
assigning to each ghost cell the value of the nearest active cell.} from
the grid at $\mathcal{R}_{\rm acc}$ (i.e.~free inflow towards
the central object) and reflection conditions at both polar boundaries. At the
outer boundary, at $\mathcal{R} = 10\, r_\mathrm{B}$, we impose the density
profile described in \eq{e3.1} and compute the pressure as $P(\theta) =
\rho(\theta)^\gamma/\gamma$, but allow free (as free-outflow condition)
evolution on both velocity components. For all the simulations we set the
fluid initially at rest (zero velocities) and uniform initial conditions
(constant density and pressure interior to the outer boundary).

We have adopted the free-outflow condition at the inner radial boundary
as we are treating the accretor as a featureless Keplerian potential. For
alternative assumptions on the accretor, e.g.~modelling it as a star,
appropriate boundary conditions should be imposed. See for example the
treatment of a hard inner boundary employed by \citet{velli94,delzanna98}
for studying time-dependent, inflow/outflow solutions in the context of
stellar atmospheres and their interaction with the interstellar medium.

All simulations were run until a stationary state was reached throughout
the domain. This was estimated by monitoring the behaviour of the mass
accretion rate $\dot{M}$, with a steady state assumed once the relative
temporal fluctuations in this parameter fell below one part in $10^{5}$. See
Figure~\ref{fig2} for an example of the time evolution of $\dot{M}$ for the
case $\gamma=4/3$. The steady state was reached at times $2000 \,t_\mathrm{B}$,
$4500 \,t_\mathrm{B}$ and $7500 \,t_\mathrm{B}$ for $\gamma = 1$, $4/3$
and $7/5$, respectively.

\subsection{Code validation}

%---------------------------------------------------------------------
\begin{figure}
\begin{center}
  \includegraphics[width=\linewidth]{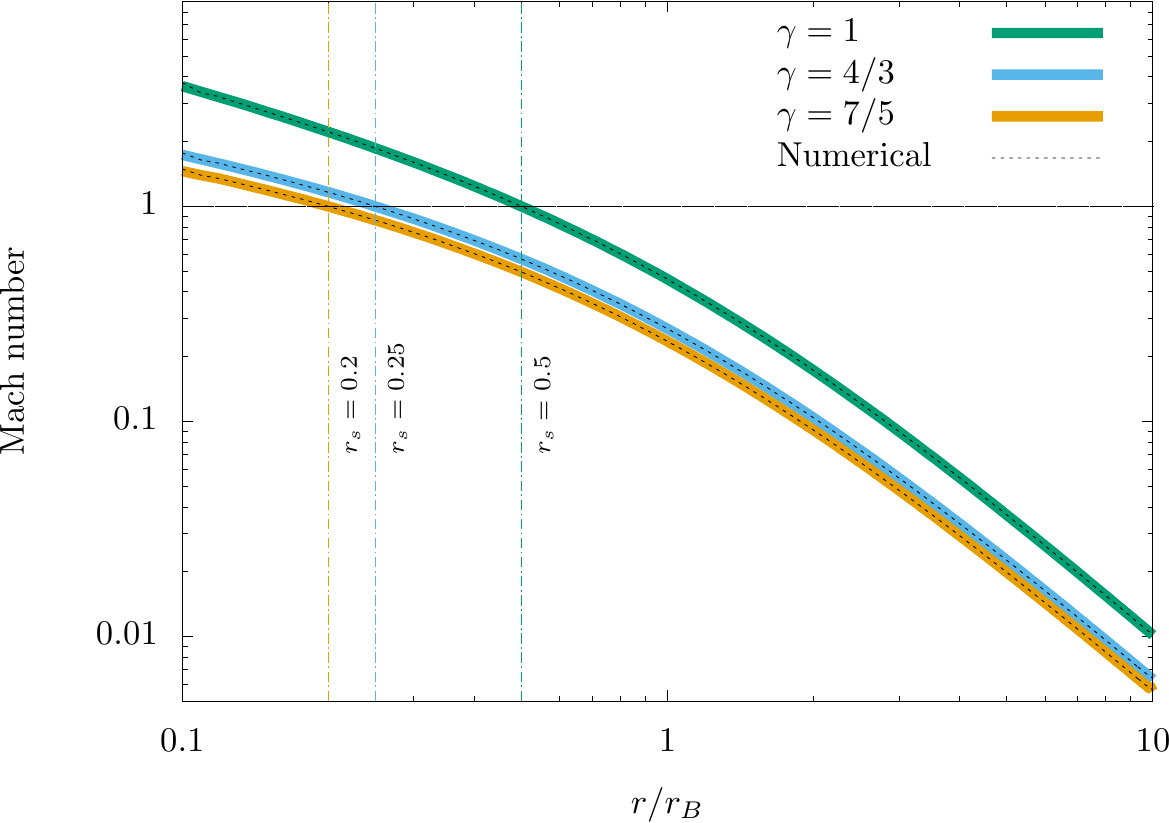}
\end{center}
\caption{Comparison of the spherically symmetric numerical simulations performed 
with {\it aztekas} and the Bondi accretion solution corresponding to the 
adiabatic indices $\gamma =1,\,4/3,\,7/5$. The figure shows the Mach number 
versus radius for the three cases. Solid, coloured lines correspond to the Bondi 
solution while dashed, black lines correspond to the numerical results. The
vertical dashed lines indicate the position of the sonic radius for
each $\gamma$.}
\label{fig3}
\end{figure}

As a validation of the {\it aztekas} code in this work, we start by considering 
$\delta = 0$, in which case spherical symmetry is recovered and the resulting 
accretion flow should coincide with the Bondi solution. In Figure~\ref{fig3} we 
show the results for the Mach number  $\mathcal{M}=v/a$ as a function of the 
radial distance to the central object, as obtained from the three simulations 
with $\gamma =1,\,4/3, \, \text{and}\, \,7/5$ and compared to the corresponding 
analytic solutions. We find a very good agreement with the Bondi solution, with 
an error of less than 1\% in the three cases~(see Table~\ref{table1b}). Further 
tests of {\it aztekas} against other analytic solutions and astrophysical 
problems can be found in~\citet{aguayo18} and~\citet{TA19}.

The treatment of the inner boundary is crucial for solving numerically
the spherical accretion problem. One important point being whether the
sonic surface is well resolved in the numerical domain or not. In the Bondi
problem, the sonic surface appears at 
\begin{equation}
r_s = \frac{5 - 3\gamma}{4} \, r_B. 
\label{eq:rs}
\end{equation}
As mentioned above, in all the simulations we placed the inner boundary at
$\mathcal{R}_\mathrm{acc} = 0.1\, r_B$ and, thus, the sonic surface is well
resolved for all values of $\gamma$.

Note that we have not included the commonly used adiabatic index for
a monoatomic gas $\gamma = 5/3$, as in this case the sonic radius vanishes
altogether. This makes a simulation which does not include artificial viscosity
unstable to numerical fluctuations appearing at the inner boundary. We prefer
not to include any such artificial effects so as to retain confidence in
that the results obtained are a robust consequence of the physics being model.

\subsection{Results}
We now present the results for numerical simulations where the assumption of
spherical symmetry for the infalling flow is relaxed, and parametrized through 
the density contrast $\delta$ as in \eq{e3.2}.

To explore the relation between total infall and total outflow in the resulting
steady state configurations, for each simulation we compute the mass
accretion rate using~\eq{e2.5}. We apply this formula at each radial grid
point and then take the average rate for all the domain. Likewise, we measure
the injected and ejected mass rates across $\mathcal{R}$ using~\eqs{e2.9} and 
\eqref{e2.10}.

A summary of all of the simulations is shown in Table~\ref{table1b} where, 
for the various adiabatic indices and equatorial to polar density contrasts 
probed, we report the different accretion, injection and ejection rates,
the ratio between ejection and injection, together with the location of the
stagnation point and the maximum velocity attained by the outflowing material.

In order to validate these results, we performed a self-convergence test of the
numerical solutions by increasing by a factor of 1.5 the resolution in both
radial and angular coordinates. We obtained a second-order convergence as
expected for a stationary shockless solution with an HRSC method \citep{LG13}.

\begin{table}
\centering
$\gamma = 1$
\begin{tabular}{ccccccccc}
\hline\hline\\[-7pt]
$\delta$ [\%] & $\frac{\dot{M}}{\dot{M}_\mathrm{B}}$ 
&$\frac{\dot{M}_\mathrm{in}}{\dot{M}_\mathrm{B}}$ & 
$\frac{\dot{M}_\mathrm{ej}}{\dot{M}_\mathrm{B}}$ &  
$\frac{\dot{M}_\mathrm{ej}}{\dot{M}_\mathrm{in}}$ & 
$\frac{\mathcal{S}}{r_\mathrm{B}}$ & 
$\frac{v_\mathrm{max}}{v_\mathrm{esc}}$ & $\frac{v_\mathrm{max}}{a}$ 
\\[4pt]\hline\\[-7pt]
0.0  & 0.99 & 0.99 & 0    & 0    & --   & --   & -- \\[2pt]
0.1  & 0.99 & 1.95 & 0.94 & 0.48 & 5.10 & 0.130 & 0.057 \\[2pt]
1.0  & 1.02 & 4.18 & 3.16 & 0.75 & 3.35 & 0.373 & 0.164 \\[2pt]
10.0 & 1.05 & 9.99 & 8.94 & 0.89 & 2.55 & 1.182 & 0.520 \\
\hline
\end{tabular}

\vspace{12pt}
$\gamma = 4/3$

\begin{tabular}{ccccccccc}
\hline\hline\\[-7pt]
$\delta$ [\%] & $\frac{\dot{M}}{\dot{M}_\mathrm{B}}$ 
&$\frac{\dot{M}_\mathrm{in}}{\dot{M}_\mathrm{B}}$ & 
$\frac{\dot{M}_\mathrm{ej}}{\dot{M}_\mathrm{B}}$ &  
$\frac{\dot{M}_\mathrm{ej}}{\dot{M}_\mathrm{in}}$ & 
$\frac{\mathcal{S}}{r_\mathrm{B}}$ & 
$\frac{v_\mathrm{max}}{v_\mathrm{esc}}$ & $\frac{v_\mathrm{max}}{a}$ 
\\[4pt]\hline\\[-7pt]
0.0  & 1.00 & 1.00  & 0     &  0    & --   & --   & -- \\[2pt]
0.1  & 1.00 & 2.55  & 1.55  &  0.61 & 4.50 & 0.130 & 0.057 \\[2pt]
1.0  & 1.01 & 5.60  & 4.59  &  0.82 & 3.29 & 0.376 & 0.166 \\[2pt]
10.0 & 1.04 & 15.44 & 14.40 &  0.93 & 2.16 & 1.186 & 0.522 \\
\hline
\end{tabular}

\vspace{12pt}
$\gamma = 7/5$
\begin{tabular}{ccccccccc}
\hline\hline\\[-7pt]
$\delta$ [\%] & $\frac{\dot{M}}{\dot{M}_\mathrm{B}}$ 
&$\frac{\dot{M}_\mathrm{in}}{\dot{M}_\mathrm{B}}$ & 
$\frac{\dot{M}_\mathrm{ej}}{\dot{M}_\mathrm{B}}$ &  
$\frac{\dot{M}_\mathrm{ej}}{\dot{M}_\mathrm{in}}$ & 
$\frac{\mathcal{S}}{r_\mathrm{B}}$ & 
$\frac{v_\mathrm{max}}{v_\mathrm{esc}}$ & $\frac{v_\mathrm{max}}{a}$ 
\\[4pt]\hline\\[-7pt]
0.0  & 0.99 &  0.99 & 0      & 0    & --   & --   & -- \\[2pt]
0.1  & 1.00 &  2.76 & 1.76   & 0.63 & 4.34 & 0.130 & 0.057 \\[2pt]
1.0  & 1.01 &  6.23 & 5.22   & 0.83 & 3.17 & 0.376 & 0.166 \\[2pt]
10.0 & 1.04 & 17.41 & 16.37  & 0.94 & 2.05 & 1.188 & 0.523 \\
\hline
\end{tabular}
\caption{Simulations summary. In this table we give the accretion, injection and ejection rates in units of the corresponding Bondi rates (equation 
\ref{Mbondi}), ratio between ejection and injection, the position of the stagnation point along the vertical axis, as 
well as the ratio of the maximum ejection velocity to the local escape velocity 
and to the local speed of sound $a$.} 
\label{table1b} 
\end{table}

\begin{figure*}
 \begin{center}
  \begin{overpic}[trim=0 65 0 0,width=0.45\linewidth]{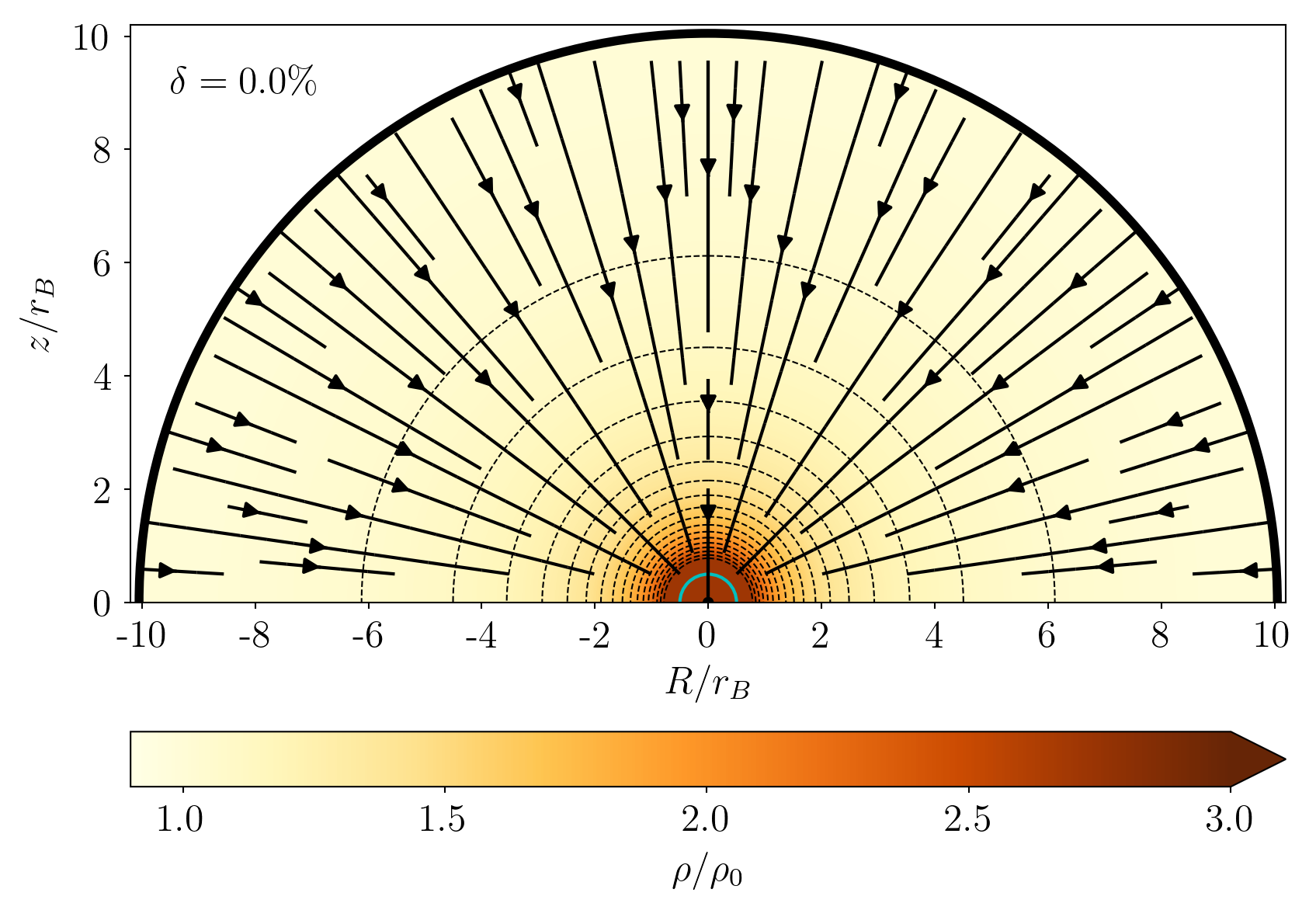}
  \end{overpic}
  \begin{overpic}[trim=0 65 0 0,width=0.45\linewidth]{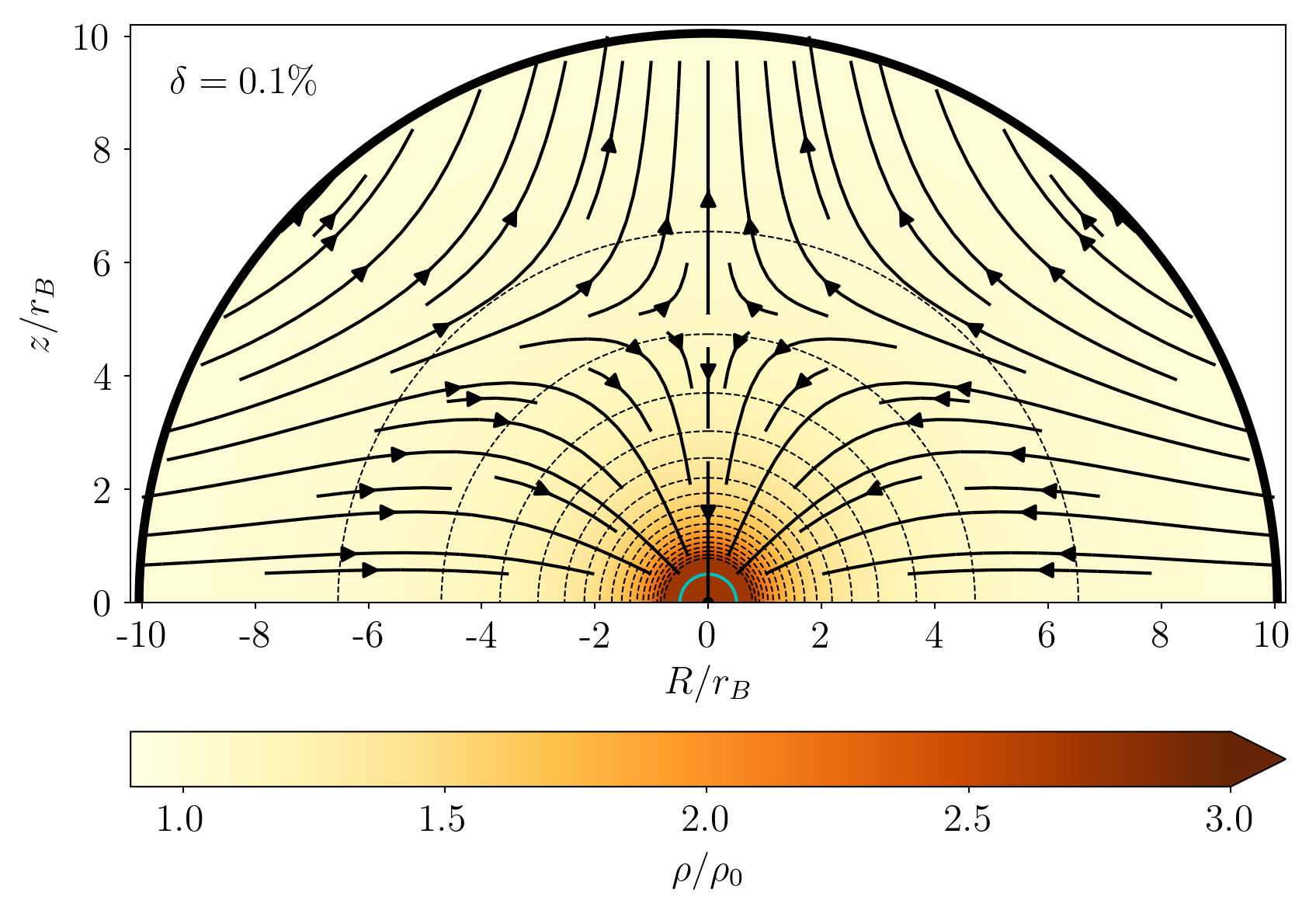}
  \end{overpic}\\
  \begin{overpic}[width=0.45\linewidth]{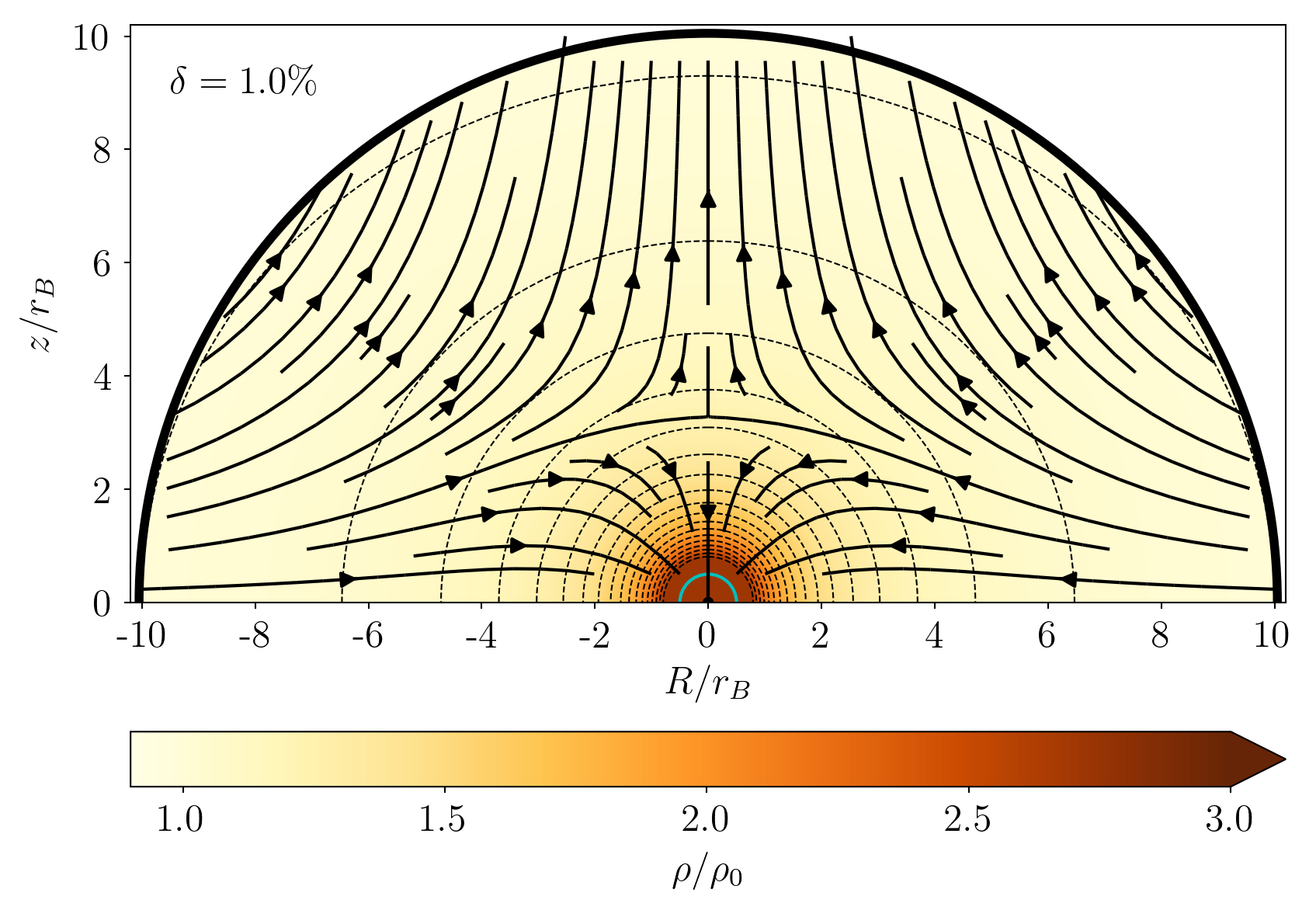}
  \end{overpic}
  \begin{overpic}[width=0.45\linewidth]{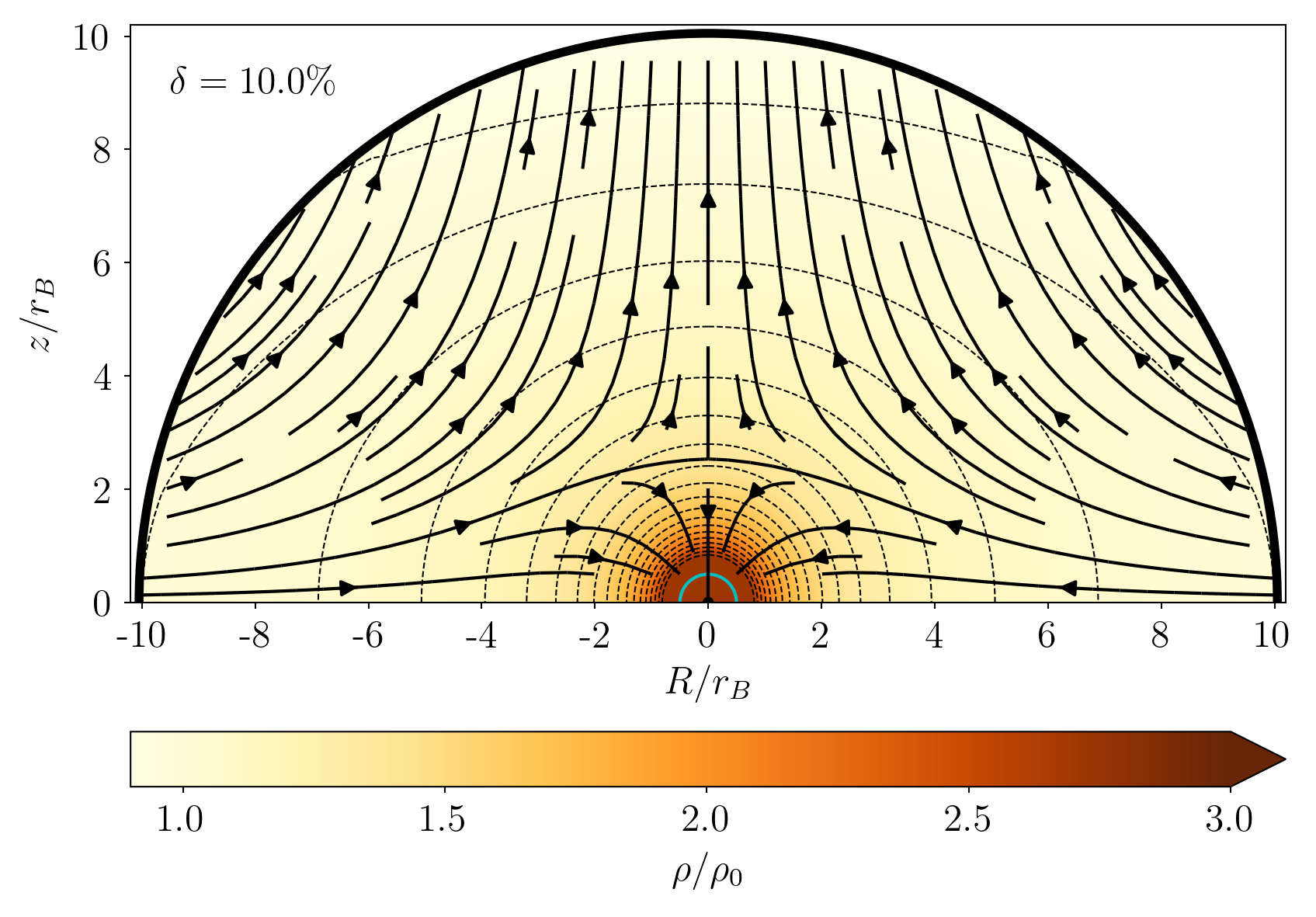}
  \end{overpic}
\end{center}
\caption{Resulting steady state flow configurations for the numerical 
simulations for a fluid with equation of state $\gamma = 1$ accreting onto a 
point mass. The different values of the density contrast $\delta$ used in 
each case are indicated on the top-left corner of each panel. The first panel 
gives the spherically symmetric case which, as expected, recovers the Bondi 
solution. As soon as $\delta\neq0$ we see a qualitative change in the flow 
morphology, with even marginal departures from sphericity resulting in polar 
outflows. As the density contrast increases, the stagnation point reaches 
deeper into the accretion flow while the outflow region expands towards the 
equator. Streamlines are shown as black solid arrows. The grey scale gives the density 
profile, with some isodensity curves shown as black dashed lines (colour version 
online). The axes correspond to the usual 
cylindrical coordinates $R = r\,\sin\theta$, $z = r\,\cos\theta$. The clear
semicircles at small radii identify the position of the Bondi sonic surface, in
all cases located within the numerical domain, which begins at $0.1 \,
r_{\mathrm{B}}$.}
\label{fig4}
 \end{figure*}

\begin{figure*}
 \begin{center}
  \begin{overpic}[trim=0 65 0 0,width=0.45\linewidth]{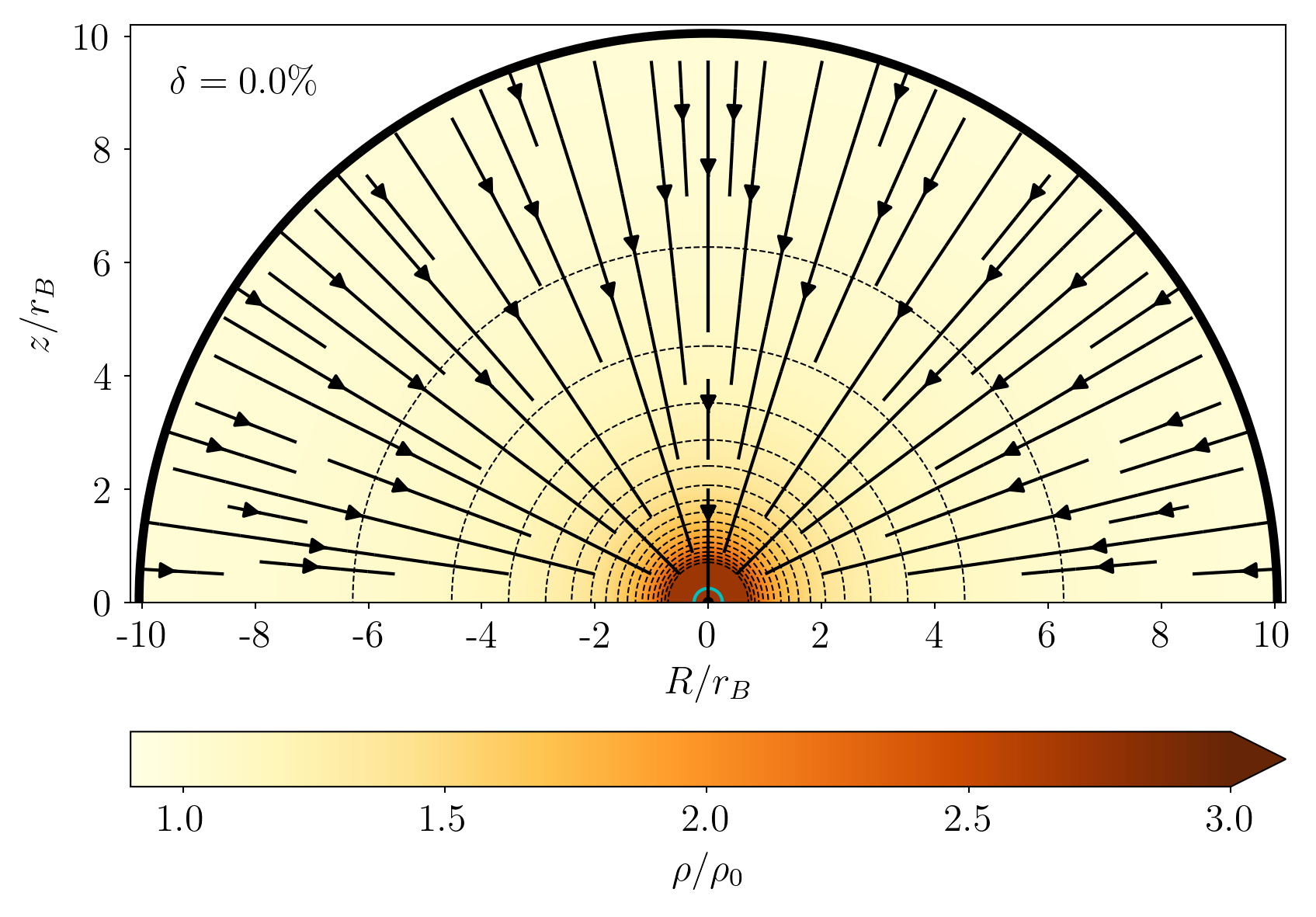}
  \end{overpic}
  \begin{overpic}[trim=0 65 0 0,width=0.45\linewidth]{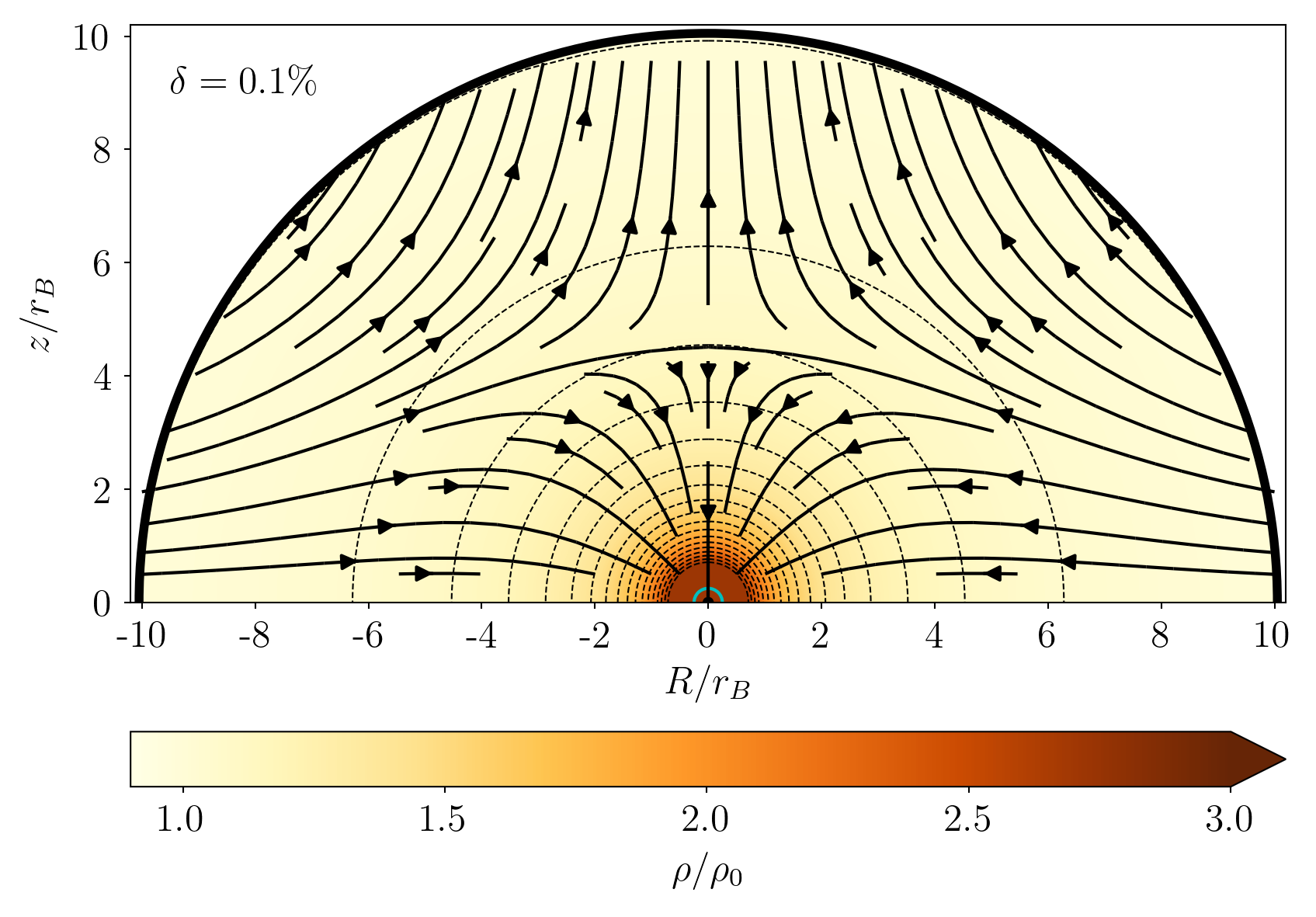}
  \end{overpic}\\
  \begin{overpic}[width=0.45\linewidth]{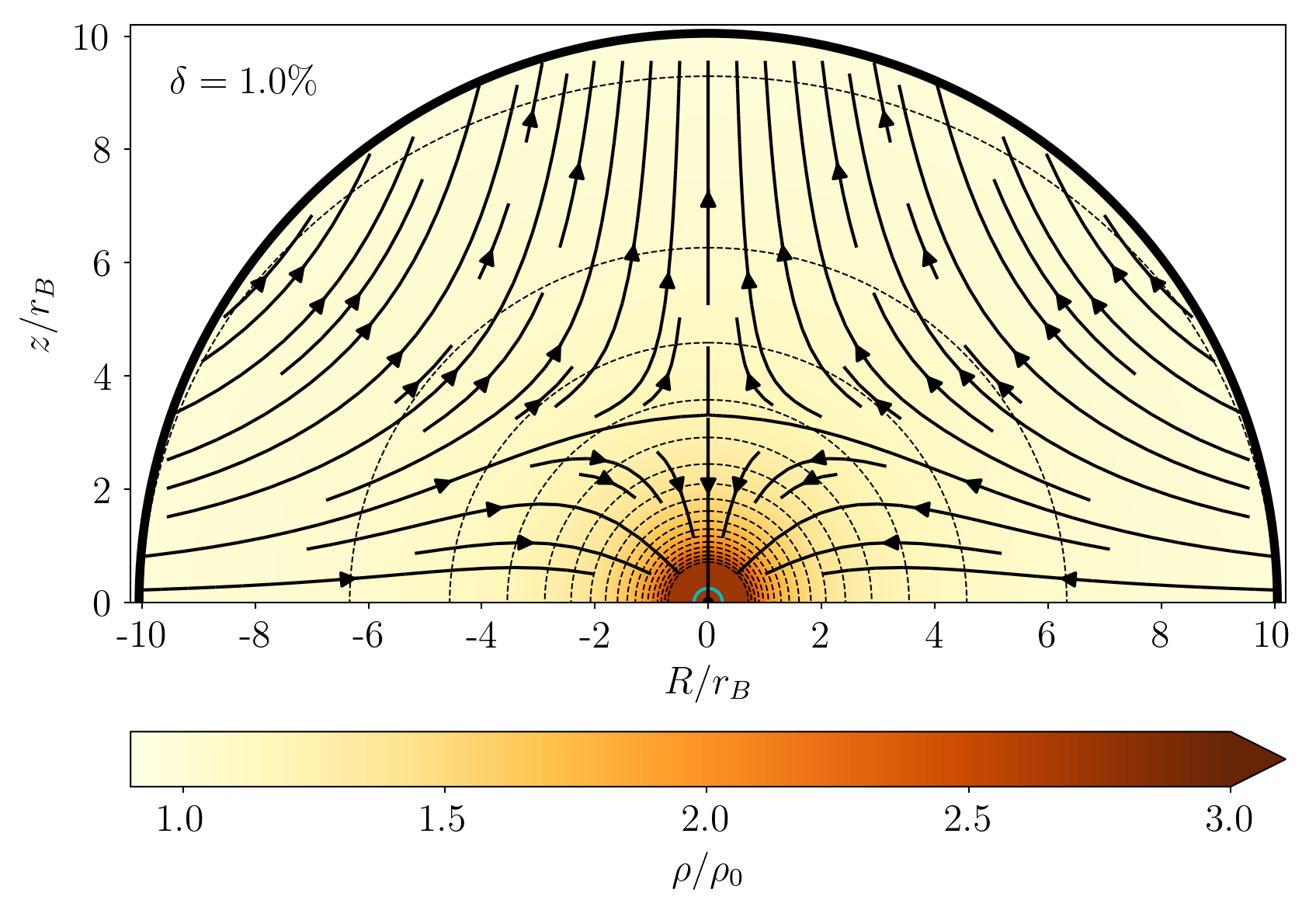}
  \end{overpic}
  \begin{overpic}[width=0.45\linewidth]{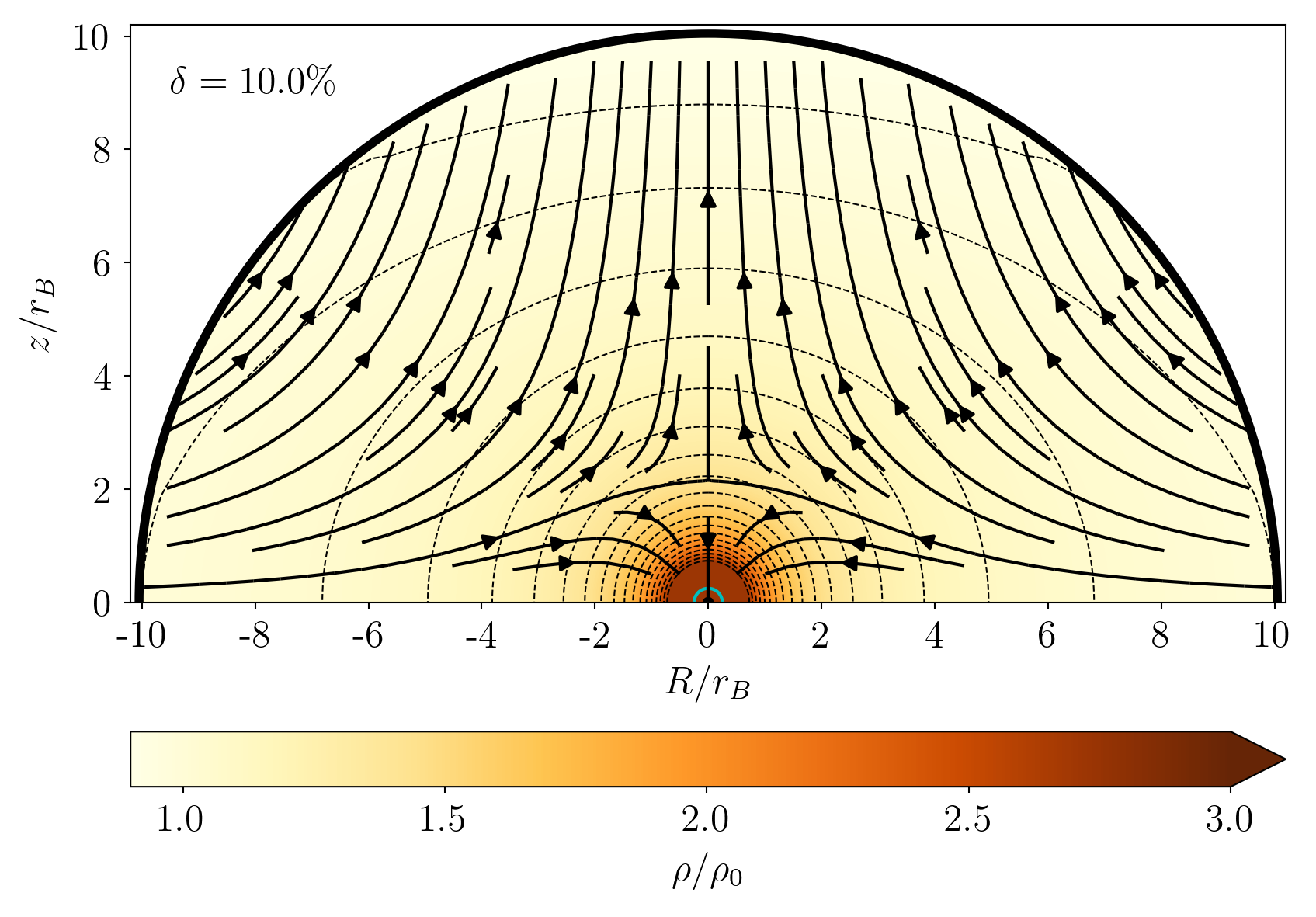}
  \end{overpic}
\end{center}
\caption{  Resulting steady state flow configurations for the numerical 
simulations with $\gamma = 4/3$. The meaning of the different lines is the same 
as in Figure~\ref{fig4}.}
\label{fig5}
\end{figure*}

\begin{figure*} 
 \begin{center}
  \begin{overpic}[trim=0 65 0 0,width=0.45\linewidth]{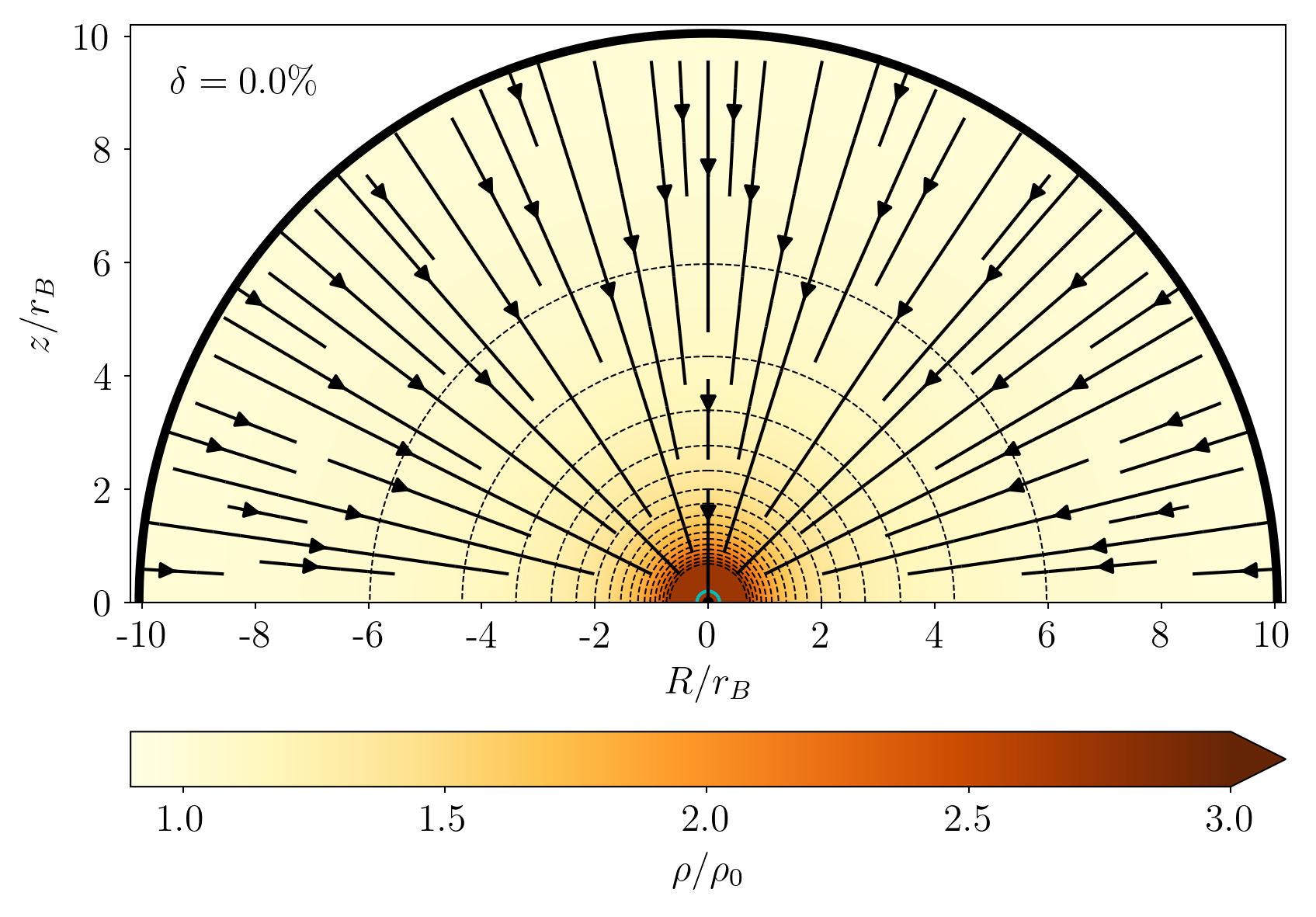}
  \end{overpic}
  \begin{overpic}[trim=0 65 0 0,width=0.45\linewidth]{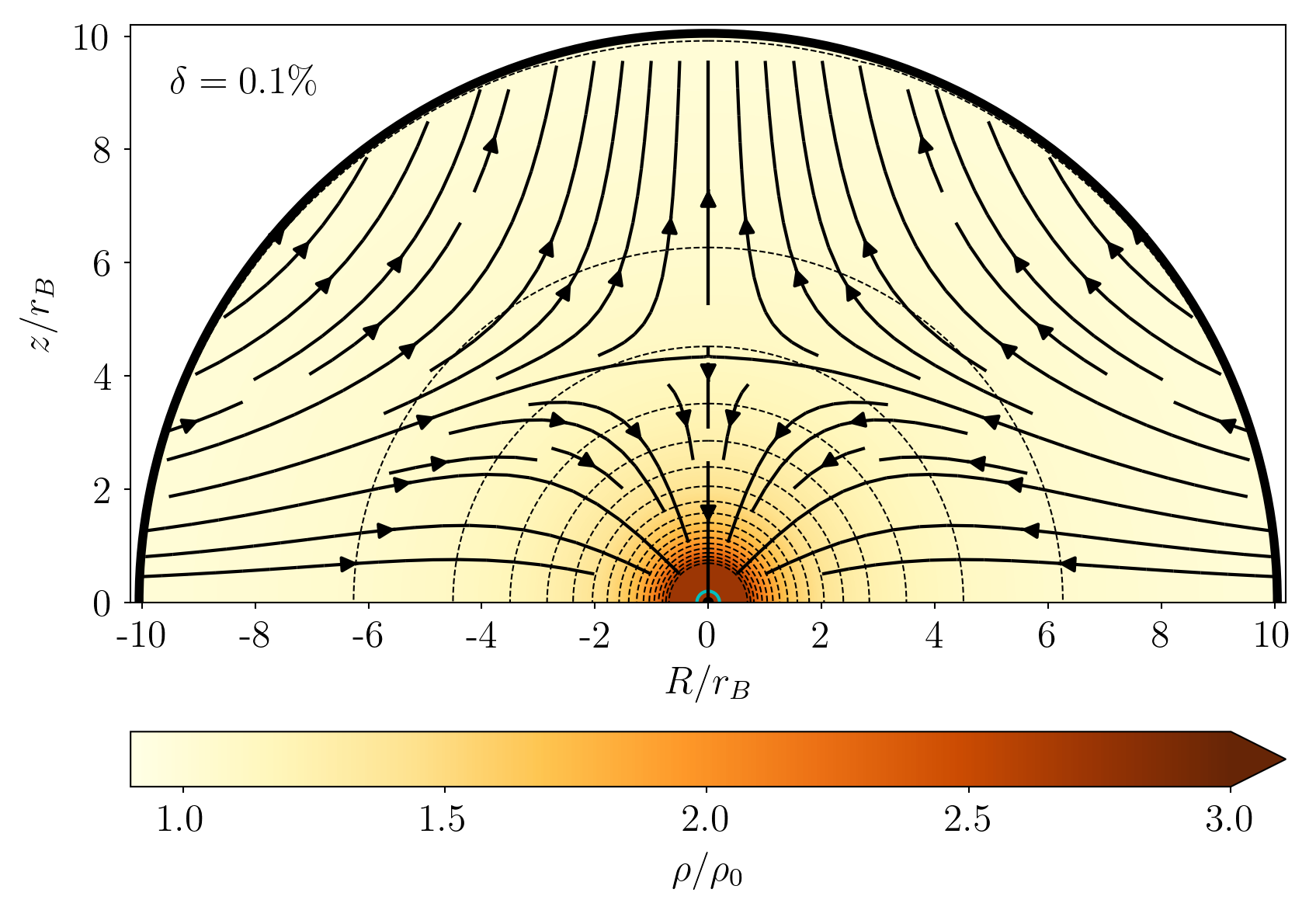}
  \end{overpic}\\
  \begin{overpic}[width=0.45\linewidth]{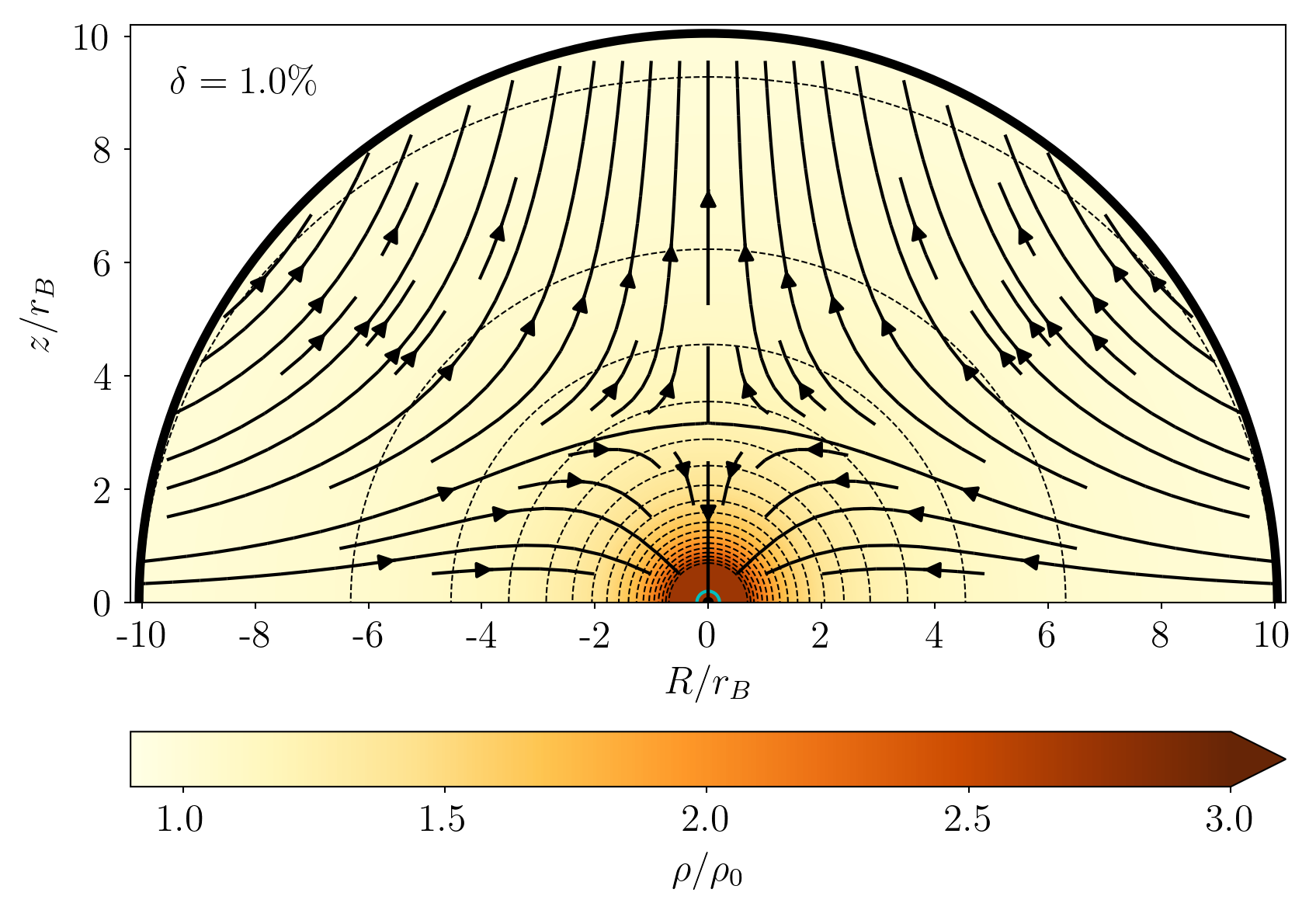}
  \end{overpic}
  \begin{overpic}[width=0.45\linewidth]{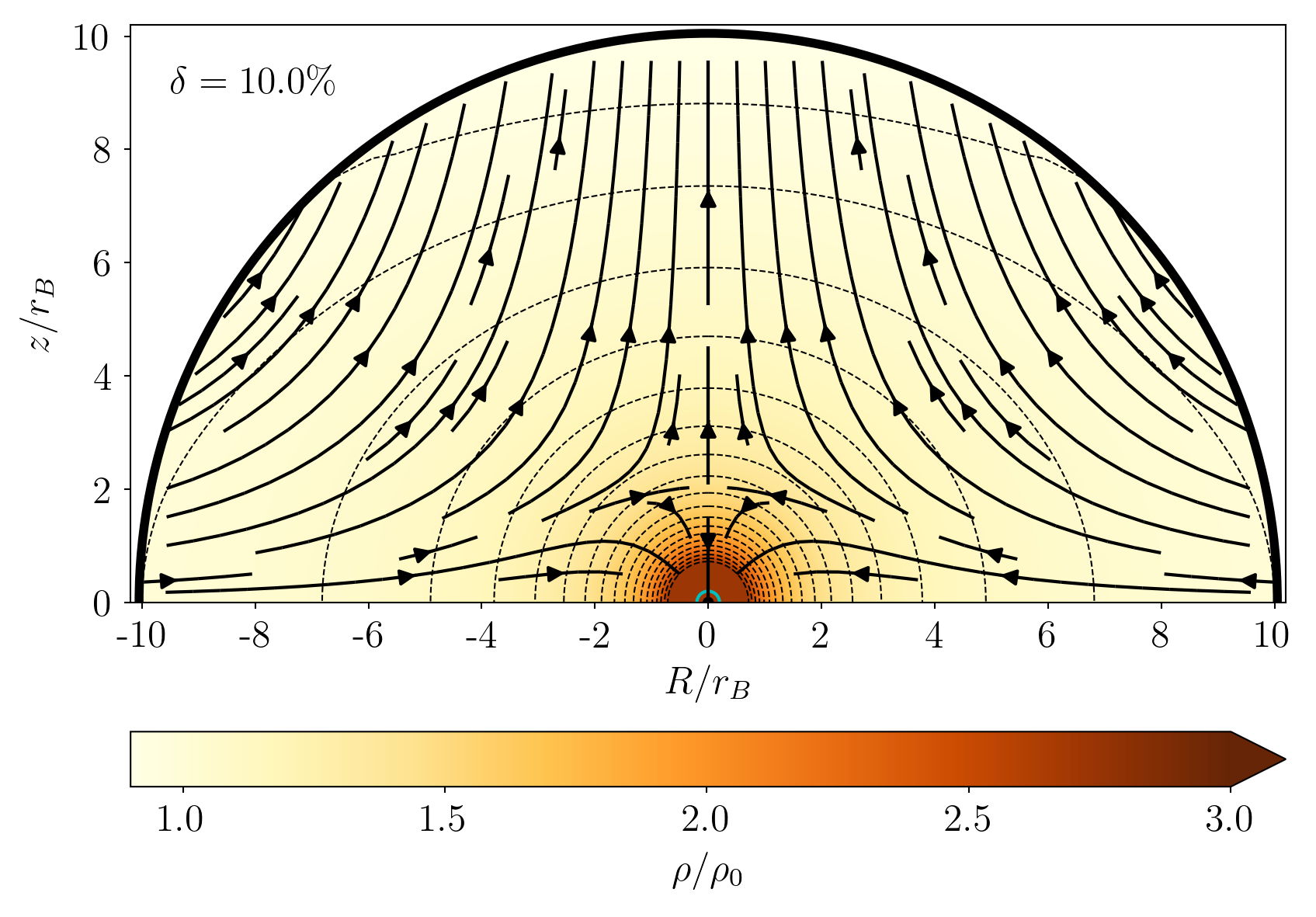}
  \end{overpic}
\end{center}
 \caption{  Resulting steady state flow configurations for the numerical 
simulations with $\gamma = 7/5$. The meaning of the different lines is the same 
as in Figure~\ref{fig4}.}
\label{fig6}
\end{figure*}

In Figure~\ref{fig4}, we show the resulting flow configurations for $\gamma 
= 1$ and the four density contrasts. The figure shows the streamlines of the 
fluid together with isocontours of the density field. Figures~\ref{fig5} and 
\ref{fig6} give the 
results for \mbox{$\gamma = 4/3$} and \mbox{$\gamma = 7/5$}, respectively.
In these figures, clear semicircles in the inner region of the numerical domain
show the locations of the sonic surfaces in the Bondi solution, which can be
seen to be resolved in all cases, as our domains start at $0.1 \,
r_{\mathrm{B}}$ in all cases.
%---------------------------------------------------------------------
%---------------------------------------------------------------------

%---------------------------------------------------------------------
%---------------------------------------------------------------------
  
%---------------------------------------------------------------------
%----------------------------------------- ----------------------------

\begin{figure} 
 \begin{center}
  \begin{overpic}[trim=0 65 0 0,width=0.95\linewidth]{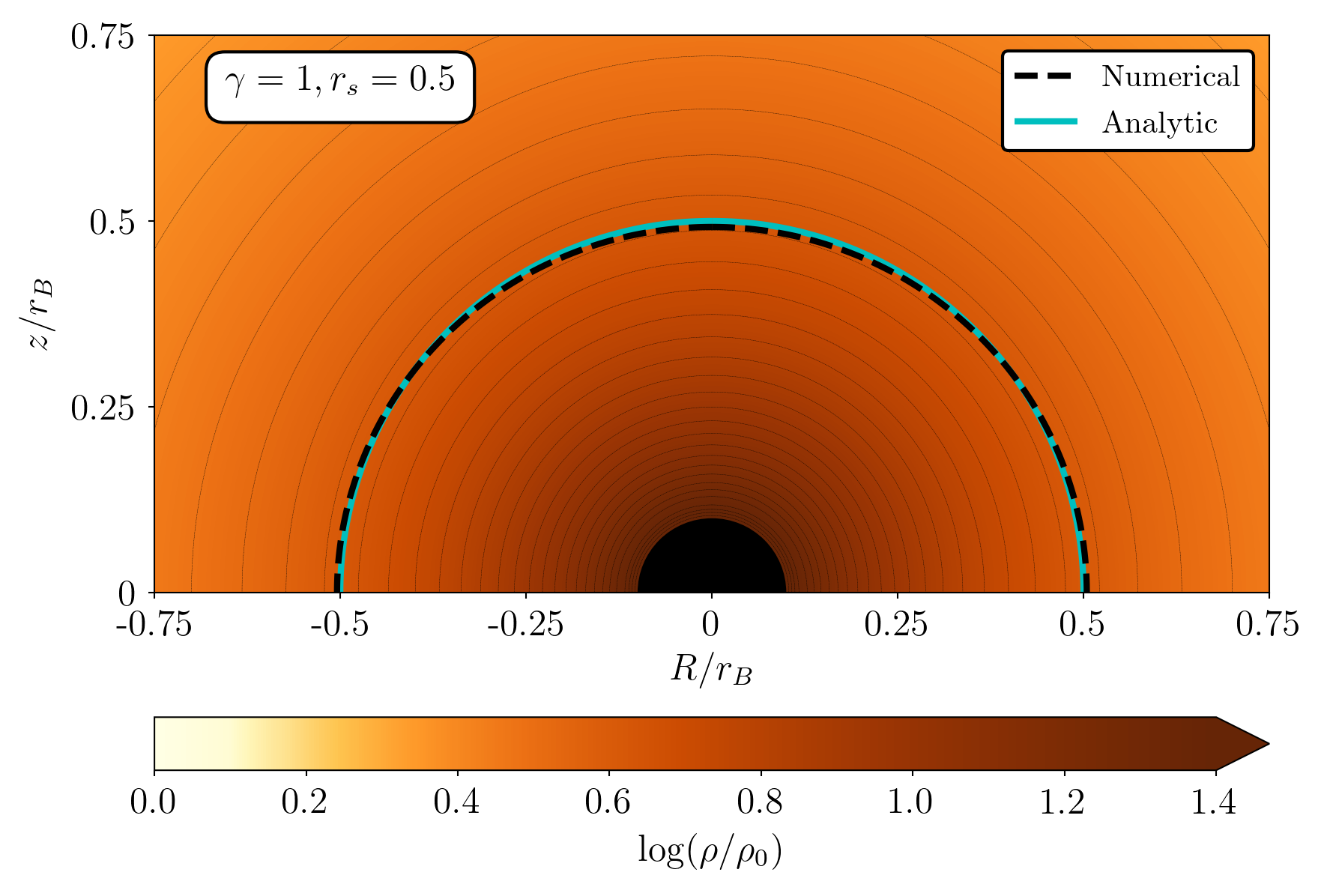}
  \end{overpic}
  \begin{overpic}[trim=0 65 0 0,width=0.95\linewidth]{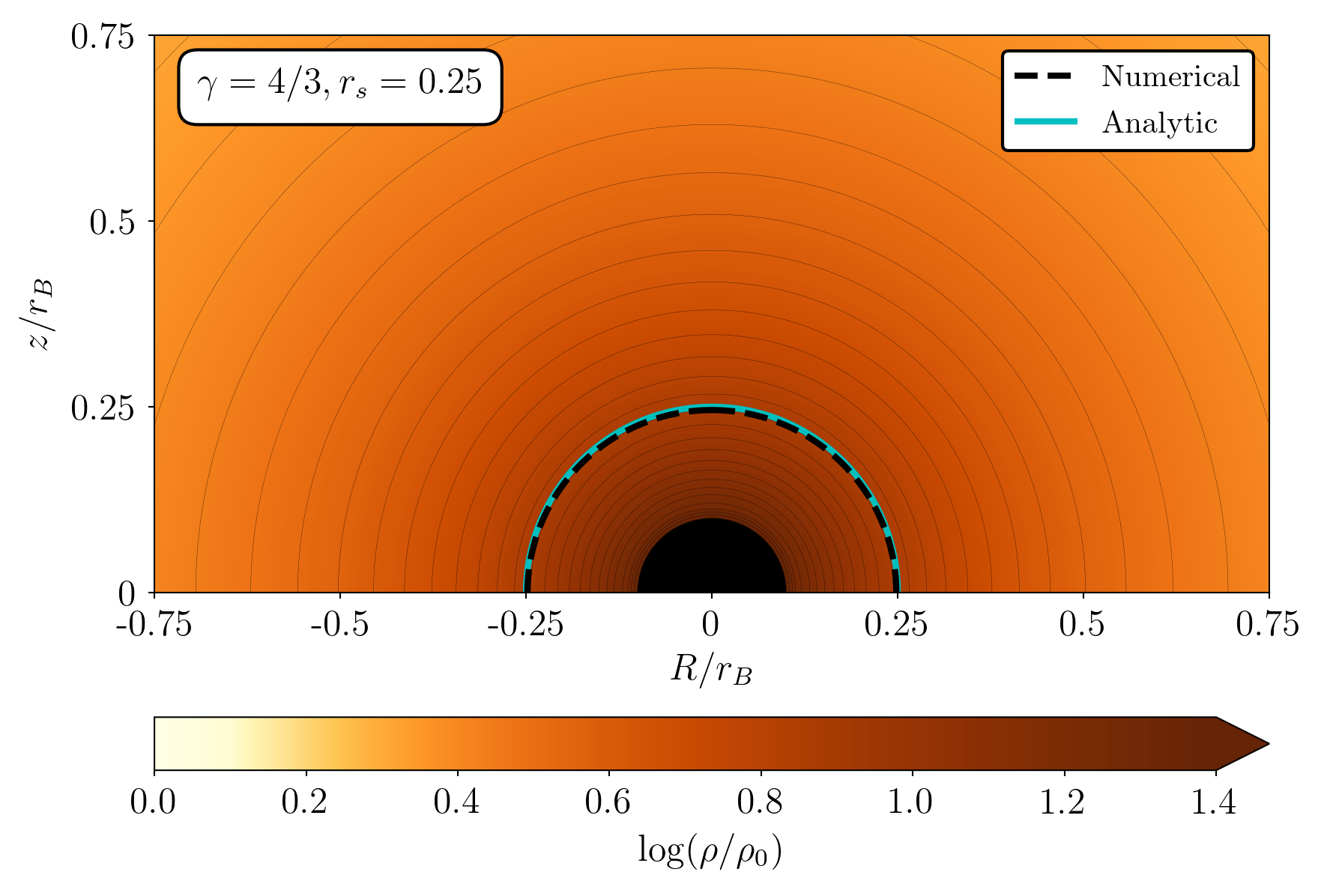}
  \end{overpic}
  \begin{overpic}[width=0.95\linewidth]{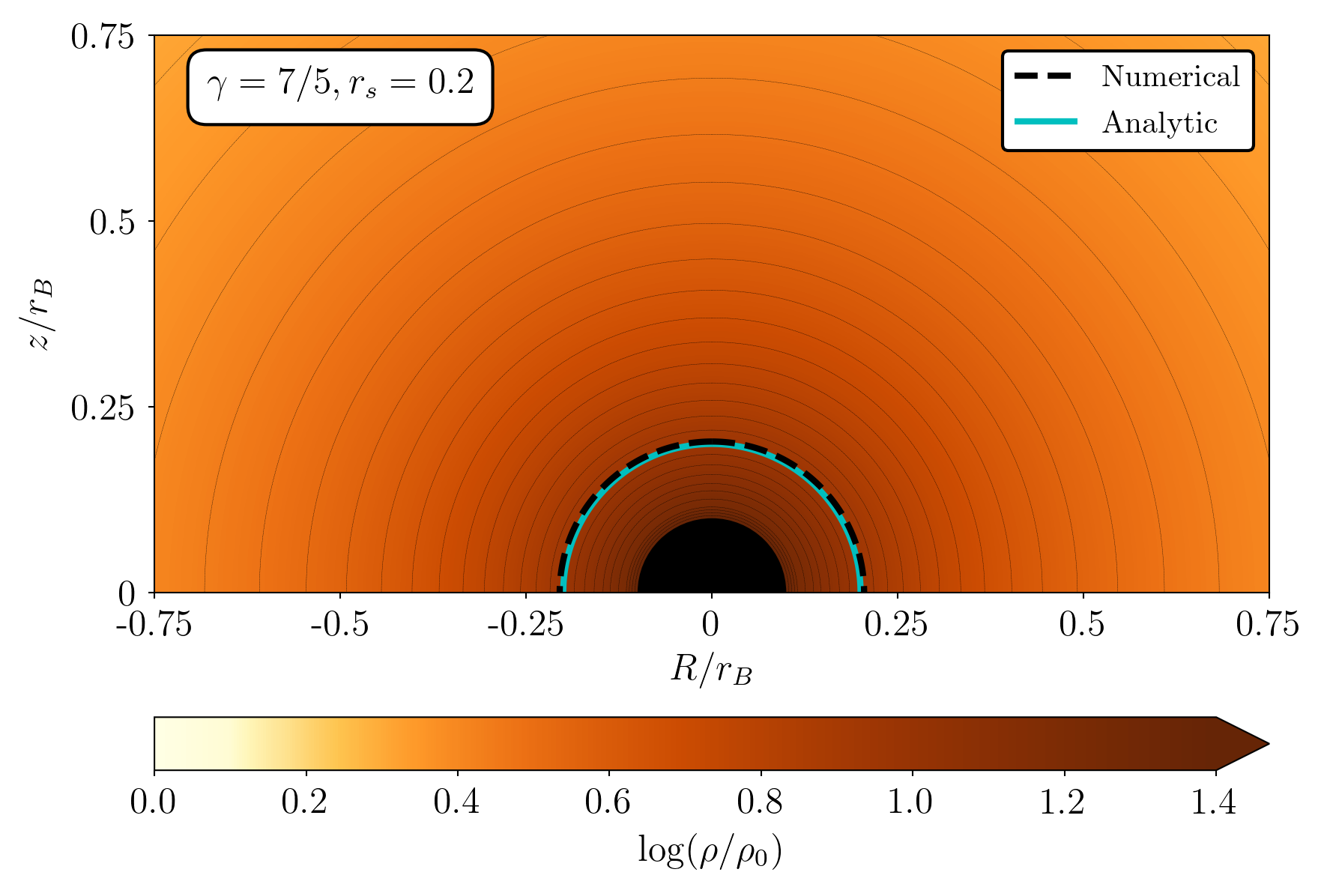}
  \end{overpic}\\
\end{center}
\caption{Close up of the inner region of the simulations with $\delta = 10\%$
for the three values of $\gamma$. In the three panels we show the location of
the sonic surface as extracted from the numerical simulations (dashed lines)
together with the one corresponding to the Bondi solution (solid line).}
\label{fig7}
\end{figure}

In Figure~\ref{fig7} we show a close up of the inner region in the case
of $\delta = 10\%$. Here we show both the analytic value of the sonic radius
(eq.~\ref{eq:rs}), as well as the one extracted from our simulations. As can
be seen, there is a very good agreement between both surfaces, even though
the numerical ones do not correspond to a spherically symmetric accretion flow.
This last illustrate the strong convergence of the solutions found towards the
Bondi solution for small radii.

Moreover, we have analysed the distribution of the entropy and of the values
of the Bernoulli constant among different streamlines. As expected for a perfect
fluid in the absence of shocks, and due to our choice of the boundary
conditions, the entropy remains constant throughout the entire numerical
domain. On the other hand, for the Bernoulli constant, although it has
a small variations from streamline to streamline, the mean value remains within
1 part in $10^4$ to the corresponding Bondi one, which agrees with the fact that
$\dot{M}\approx\dot{M}_{\mathrm{B}}$.

%---------------------------------------------------------------------
%---------------------------------------------------------------------
 \setcounter{equation}{0}

\section{Discussion}
\label{S4}

Based on the hydrodynamical numerical simulations presented in the previous
section, we have explored the consequences of breaking spherical symmetry by
imposing an axisymmetric, polar density contrast $\delta$ in the accretion
flow. Naturally, when $\delta = 0$ the spherically symmetric Bondi solution
is recovered, as can be seen from Figure~\ref{fig3} and the top-left panel
on Figures~\ref{fig4}-\ref{fig6}. However, as soon as there is a non-zero
density contrast between the equatorial plane and the polar regions, the flow
morphology changes qualitatively into an inflow/outflow configuration. The
onset of even a quite marginal density contrast of $\delta = 0.1\%$ results
in the appearance of flow patterns highly resembling the analytic toy model
discussed in Section~\ref{S2} as well as the inflow/outflow configurations
of the perturbative results of \cite{hernandez14} for an isothermal model.

As can be seen from the results reported in Table~\ref{table1b},  as we
take progressively larger values for the density contrast, the magnitude of
the ejection velocities increases, and the position of the stagnation point
moves towards smaller radii as the outflow region expands to occupy a larger
fraction of the simulation domain.  Note from Figures~\ref{fig4}-\ref{fig6}
that for $\delta=10\%$, the density contours appear slightly oblate, with
the departure from spherical symmetry becoming more apparent. Moreover,
for this same density contrast we obtain ejection velocities larger than the
local escape velocity, while, at the same time, the ratio of mass ejection
and mass injection rates reaches values as high as 94$\%$.

We have presented simulation results for three different values of the
adiabatic index: $\gamma=1$, 4/3 and 7/5. Although we also explored
intermediate values of $\gamma$, we do not include any further results,
as they are all very similar and show a continuous progression between the
cases presented. Indeed, we see that the three cases already considered are
qualitatively very similar, as could have been anticipated from the qualitative
similarities of the analytic solution of the highly simplified incompressible
model presented in Section~\ref{S2} and the slightly more realistic streamlines
of the isothermal perturbative solutions of \cite{hernandez14}. It is
clear that breaking spherical symmetry with a polar density gradient in
hydrodynamical accretion models leads to the same qualitative solutions with
a morphology as shown in Figures \ref{fig4}-\ref{fig6}.

\begin{figure}
\begin{center}
  \includegraphics[width=\linewidth]{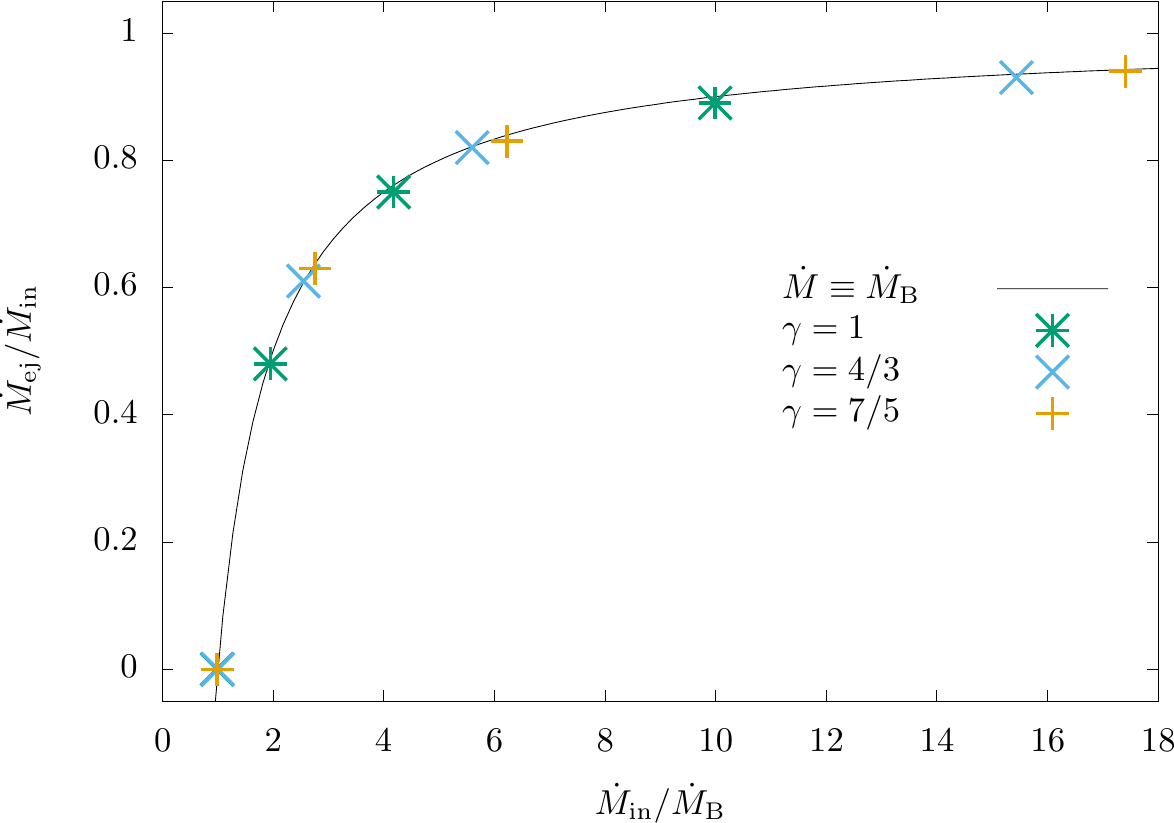}
\end{center}
\caption{Ratio between the ejected and injected mass rates versus the injected
mass rate in units of $\dot{M}_\mathrm{B}$ (see equation \ref{Mbondi}) 
for all of the simulations reported in this work.
The solid line represents the case where the accretion rate onto the central
object $\dot{M}$ is equal to the Bondi mass accretion rate.}
\label{fig8}
\end{figure}

In Figure~\ref{fig8} we show, for all of the simulations
reported in this work, the ratio of the ejected and injected
mass rates ($\dot{M_\mathrm{ej}}/\dot{M_\mathrm{in}}$) versus
the injected mass rate in units of the corresponding Bondi value
($\dot{M_\mathrm{in}}/\dot{M_\mathrm{B}}$). The solid line represents the
points where the injection and ejection of material are such that the total
mass accretion rate equals the Bondi value, i.e.~
\begin{equation}
\dot{M} = \dot{M}_{\rm in} - \dot{M}_{\rm ej} \equiv \dot{M}_\mathrm{B}.
\end{equation}
From this figure we see that for 
$\dot{M_\mathrm{in}}/\dot{M_\mathrm{B}}=1$, the ejected mass is zero, while as  
$\dot{M_\mathrm{in}}/\dot{M_\mathrm{B}}$ increases,  the ratio 
$\dot{M_\mathrm{ej}}/\dot{M_\mathrm{in}}$ tends to unity. It is remarkable that 
this dependency between injection and accretion holds quite accurately across
the adiabatic index range sampled. Differences between $\dot{M}$ and
$\dot{M_\mathrm{B}}$  remain beyond numerical resolution, but in all cases below
a $5\%$ level.  Thus, regardless of how large the injected mass rate is at the
outer boundary, the central object only accretes at a maximum critical rate,
closely corresponding to the Bondi mass accretion rate. This justifies our
choice in \eq{Mbondi} for the mass accretion rate of the analytic model of
choked accretion.

The bipolar boundary density distribution adopted in \eq{e3.1} is not intended
as a physical model corresponding to any specific situation, but rather as a
convenient first order parametrization of departure from spherical symmetry.
However, infall processes originating from accretion disc phenomena will quite
probably be characterized by axisymmetric density profiles denser towards the
equator than the poles. The details of the flow patterns resulting from
particular situations will surely differ from the results presented here,
although the clear convergence towards spherical accretion seen at small radii
implies robustness in our conclusions with respect to this point. Indeed, the
overall flow patterns obtained persist even when changing the particular
parametrization used. Other similar density profiles were explored, obtaining
consistent results, although very distinct boundary conditions may very well
lead to substantially different solutions. 

%---------------------------------------------------------------------
%---------------------------------------------------------------------

Although the imposed geometry of the inflow boundary conditions
necessarily leads to final configurations sharing the same symmetry, the
resulting inflow/outflow morphology is by no means the only configuration
sharing this symmetry, e.g. a pure infall configuration with oblate isodensity
contours would have also satisfied the symmetry conditions alone. Further, the
resulting limit on the accretion rate saturating close to the Bondi value
is in no way evident simply from the assumption of a polar density gradient
on the boundary conditions.

Let us explore now the viability of the choked accretion model to contribute
towards the launching of a jet in different astrophysical settings. For
this model we have neglected the effects of fluid rotation, which is a valid
approximation just for the innermost part of an accretion disc where, through
viscous transport mechanisms \citep[e.g][]{balbus}, the disc material has
lost most of its angular momentum. It is then natural to ask for the whole
choked accretion machinery to fit within the inner walls of the disc. In other
words, we require $\mathcal{S}$, the characteristic length scale of the choked
accretion model, to be comparable to the physical size of this inner region.

Our simulation results show that $\mathcal{S}$ sinks deeper into the
central region as we consider both larger density contrasts $\delta$ and
larger adiabatic indices $\gamma$. We note, however, that for the parameters
explored in this work $\mathcal{S}$ is always larger than the Bondi radius
$r_B = \G M/a^2_\infty$. Assuming that this result holds in general, and
considering an ideal gas composed of monoatomic hydrogen, we can write
\begin{equation}
a_\infty = \sqrt{ \frac{\gamma\,k_B\,T_\infty}{m_H} } \simeq 3\times10^{5} 
\left(\frac{T}{10^3\mathrm{K}}\right)^{1/2}\mathrm{cm}/\mathrm{s},
\end{equation}
and, consequently, we have
\begin{equation}
\mathcal{S} \gtrsim 100 \left(\frac{M}{M_\odot}\right)
\left(\frac{T}{10^3\mathrm{K}}\right)^{-1}\mathrm{au},
\label{stag1}
\end{equation}
or, in units of the gravitational radius $r_g = \G M /\cc^2$
\begin{equation}
\mathcal{S} \gtrsim 5\times10^6 \left(\frac{T}{10^6\mathrm{K}}\right)^{-1}\,r_g.
\label{stag2}
\end{equation}

These expressions should be considered as upper limits for
$\mathcal{S}$. Additional physical ingredients that have been left out
from this simple model can contribute to reduce this characteristic length
scale. For example, we have neglected the necessary presence of radiation
and magnetic fields in the system, both of which will result in an enhanced
effective temperature, which will in turn result in smaller values of
$\mathcal{S}$.

Let us consider the accretion discs behind X-ray binaries or
AGNs. For these systems the inner border of the disc is expected to
have a radius of the order of $1 - 10\,r_g$  and a temperature of around
$T\sim10^6\,\mathrm{K}$~\citep{agn,xray}. From \eq{stag2} we have $\mathcal{S}
\gtrsim 10^6 \,r_g$ and we can then conclude that these systems are too cold
for choked accretion on its own to be responsible for the ejection.

On the other hand, for a Keplerian protoplanetary disc around a YSO we have
inner radii in the order of $0.1 - 1.0\,\mathrm{au}$ and temperatures that can
reach up to $10^3\,\mathrm{K}$ \citep{proto}. In this case from \eq{stag1}
we obtain $\mathcal{S} \gtrsim 100 \,\mathrm{au}$, which is two to three
orders of magnitude larger than the presumed physical size of the system.
Nevertheless, choked accretion might contribute to the ejection of molecular
outflows associated to YSOs \citep{bachiller96}, as these outflows are
driven by infalling matter and originate from regions farther than several
$100 \,\mathrm{au}$ from the central accretor.

Lastly, if we consider the central progenitors of long GRBs within the
so-called collapsar scenario \citep{woosley}, we have again inner radii
of the order of $1 - 10\,r_g$ but temperatures that can now reach up to
$10^{11}\,\mathrm{K}$. From \eq{stag2} for this temperature we obtain
$\mathcal{S} \gtrsim 50 \,r_g$, which is now within the expected order of
magnitude for the mechanism presented to become relevant.

From these examples, it might seem that the length scale $\mathcal{S}$ on which
the choked accretion mechanism operates is too large to contribute towards
the ejection mechanism behind most astrophysical jets. It is important to
keep in mind, however, that relaxing some of the simplifying assumptions behind
the present model will necessarily lead to smaller values of $\mathcal{S}$. In
addition to the already mentioned role played by radiation and magnetic fields,
we have also seen a clear trend for diminishing values of $\mathcal{S}$
as the equatorial to polar density contrast increases. Hence, we can expect
the applicability of the choked accretion mechanism to increase once larger
density contrasts are considered.

In connection to the previous point, a non-zero angular momentum will naturally
contribute to increase the density contrast between the material in the disc
with respect to the polar regions \citep[see e.g.][]{tejeda1}. We were not able
to explore this regime in this work since from our numerical experiments we
have found that taking larger values of $\delta$ or outer boundaries smaller
than $r_\mathrm{B}$ lead to rather unstable, highly dynamic accretion flows
that do not seem to relax to steady state configurations. We plan, however,
to study both the effect of rotation and the dynamic regime resulting from
steeper density gradients in future work.

There are additional physical processes, particularly non-adiabatic effects
such as radiative cooling, heat transfer and shock formation, that can in
principle modify our results. However, it is not clear {\it a priori} whether
the inclusion of these effects will enhance or hamper the applicability of
the choked accretion mechanism. We leave addressing these important points
as the focus of future investigations, whereas the current simple model can
be seen as presenting an underlying phenomenon which might be an important
factor within a more general scheme.

\setcounter{equation}{0}
\section{Conclusions}
\label{S5}

Through simple analytic considerations for an idealized incompressible fluid,
and axisymmetric numerical simulations for hydrodynamical accretion in a
central Newtonian potential spanning a range of adiabatic indices, we have
shown that large-scale, small amplitude departures from spherical symmetry,
where the polar regions are under dense with respect to the equatorial plane,
result in substantial qualitative and quantitative modifications in the
resulting flow morphology.

Although the details depend on the particular choice of physical parameters,
for the continuous bipolar boundary conditions explored, the resulting flow
patterns are no longer the classical radial accretion ones,  but have a
consistent morphology comprising a central quasi-spherical accretion  region,
and an outer zone of equatorial infall and polar outflows.

We have seen that as the density contrast increases, a larger fraction of the
mass injection rate is reversed and expelled along a bipolar outflow. Moreover,
the maximum velocity attained by this ejected material can reach values
larger than the local escape velocity.

As the accretion flow tends towards spherical symmetry at small radii, we
find the interesting result that the total mass accretion rate is limited at
a few percent above the Bondi value of the corresponding spherically symmetric
case. Thus, hydrodynamical accretion towards central objects appears to choke
at a maximum value, with any extra input material fuelling the bipolar outflow.

In conclusion, we have presented compelling evidence for the existence of
a choked accretion phenomenon in hydrodynamical flows onto gravitating
objects. Furthermore, we have shown that choked accretion works as a
hydrodynamical mechanism for ejecting axisymmetric outflows, thus constituting
a transition bridge between purely radial accretion flows and jet-generating
systems.

\section*{Acknowledgements}

We thank Sergio Mendoza, John Miller, Olivier Sarbach, Francisco Guzm\'an,
Diego L\'opez-C\'amara, Fabio de Colle and Jorge Cantó for useful discussions
and comments on the manuscript. The authors also acknowledge the constructive
criticism from an anonymous referee. This work was supported by DGAPA-UNAM
(IN112616 and IN112019) and CONACyT (CB-2014-01 No.~240512; No.~290941;
No.~291113) grants. AAO and ET acknowledge economic support from CONACyT
(788898, 673583). XH acknowledges support from DGAPA-UNAM PAPIIT IN104517
and CONACyT.

\bibliographystyle{mn2e}

\bibliography{references}

\label{lastpage}

\end{document}